\documentstyle[preprint,prb,aps,epsfig]{revtex}

\begin{document}
\title{A Monte Carlo Test of the Fisher-Nakanishi-Scaling Theory for the Capillary
Condensation Critical Point}
\author{Oliver Dillmann$^{\text{1}}$, Wolfhard Janke$^{\text{2}}$, Marcus M\"{u}ller$%
^{\text{1}}$ and Kurt Binder$^{\text{1}}$}
\address{$^{\text{(1)}}$ Institut f\"{u}r Physik, Johannes-Gutenberg-Universit\"{a}t\\
Mainz, D-55099 Mainz, Staudinger Weg 7, Germany\\
$^{\text{(2)}}$ Institut f\"{u}r Theoretische Physik, Universit\"{a}t\\
Leipzig, D-04109 Leipzig, Augustusplatz 10/11, Germany.\\
email: Wolfhard.Janke@itp.uni-leipzig.de; Kurt.Binder@uni-mainz.de}
\maketitle

\begin{abstract}
Extending the Swendsen-Wang cluster algorithm to include both bulk ($H$) and
surface fields ($H_1$) in $L\times L\times D$ Ising films of
thickness $D$ and two free $L\times L$ surfaces, a Monte Carlo study of the
capillary condensation critical point of the model is presented. Applying a
finite-size scaling analysis where the lateral linear dimension $L$ is
varied over a wide range, the critical temperature $T_{c}(D)$ and the
associated critical field $H_{c}(D)$ are estimated for $4\leq D\leq 32$
lattice spacings, for a choice of the surface field $H_1$ small
enough that the dependence of $H_{c}(D)$ on $H_1$ is still linear.
It is shown that the results are consistent with the power laws predicted by
Fisher and\ Nakanishi [M.E. Fisher and H. Nakanishi, J. Chem. Phys. 75, 5857
(1981)], namely $T_{c}(\infty )-T_{c}(D)\propto D^{-1/\nu }$, $%
H_{c}(D)\propto D^{-(\Delta -\Delta _{1})/\nu }$, where $\nu $ is the bulk
correlation length exponent of the three-dimensional Ising model, and $%
\Delta $, $\Delta _{1}$ are the corresponding ``gap exponents'' associated
with bulk and surface fields, respectively. As expected, the order parameter
of the thin film near its critical point exhibits critical behavior
compatible with the universality class of the two-dimensional Ising model.
\end{abstract}

\section{Introduction}

The application prospects of nanoscale technology have created a fresh
interest in the behavior of both simple fluids and complex fluids confined
in pores or in a thin film geometry in layers confined by parallel walls 
\cite{1,2,3,4,5}. However, a prerequisite for the clarification of pattern
formation \cite{2,3} and dynamics \cite{4,5} is a good understanding of the
interplay between bulk and surface effects on thermodynamics and phase
behavior in this finite-size geometry \cite{6,7,8}: although theoretical
aspects of phase transitions and critical phenomena in confined geometry
have been considered since a long time \cite
{9,10,11,12,13,14,15,16,17,18,19,20,21,22,23,24,25,26,27,28}, this is still
a topic of active current research \cite{29,30,31,32,32a} even for one of the
most well-known phenomena, namely ``capillary condensation'' \cite{33,34}.
By ``capillary condensation'' one means the finding, already discovered in
the 19th century \cite{33}, that in a capillary the condensation of a gas
occurs already at a lower pressure $p$ than the coexistence pressure $p_{%
\text{coex}}$ necessary to induce condensation in the bulk. Qualitatively,
this shift of the transition can be attributed to the interaction of the
fluid molecules with the attractive walls of the capillary. Although
confinement effects on fluids and their phase transitions have been studied
experimentally since a long time as well \cite
{35,36,37,38,39,40,41,42,43,44,45,46,47,48}, a quantitative characterization
of the shift of the capillary condensation critical point remains a
challenge. While for temperatures $T$ below the critical temperature $%
T_{c}(D)$ of the thin film of thickness $D$ the chemical potential at the
condensation transition $\mu _{c}(D)$ is shifted relative to its bulk value
simply as $\mu _{c}(D)-\mu _{c}(\infty )\propto D^{-1}$ (``Kelvin
equation'') \cite{27}, for large enough $D$, Fisher and Nakanishi \cite{16}
predicted a completely different power law for the corresponding shift at
the critical temperature itself, namely 
\begin{equation}
\mu _{c}(D)-\mu _{c}(\infty )\propto D^{-(\Delta -\Delta _{1})/\nu },\quad
T=T_{c}(D)  \label{eq1}
\end{equation}
for weak surface forces. In Eq.~(\ref{eq1}), critical exponents of the
three-dimensional Ising model universality class (that encompasses
criticality of gas-fluid critical points or the related unmixing transitions
in binary mixtures, etc.) enter, namely the correlation length exponent \cite
{49} $\nu \approx 0.63$ and the ``gap exponent'' $\Delta = \gamma + \beta
\approx 1.56$ and the corresponding exponent for a free surface \cite
{19,50,51,52} $\Delta _{1}\approx 0.46-0.48$. Also for the shift of $T_{c}$
a similar power law holds, 
\begin{equation}
T_{c}(\infty )-T_{c}(D)\propto D^{-1/\nu },  \label{eq2}
\end{equation}
which is the same relation as is familiar from standard finite-size scaling 
\cite{10,18,53,54} for the shift of $T_{c}$ in films with ``neutral walls''
\{i.e., no surface field preferring one of the phases coexisting for $%
T<T_{c}(D)$ at $\mu _{c}(D)\equiv \mu _{c}(\infty )$ act\} or in films with
periodic boundary conditions where surface effects are a priori absent.
While for the latter systems Eq.~(\ref{eq2}) has been studied by various
methods \cite{13,15,55,56,57,58,59}, Eq.~(\ref{eq1}) to our knowledge has not
yet been tested by Monte Carlo simulations. In previous work \cite{26,28,29}%
, tests of the Kelvin equation and corrections to it \cite{60} have been
carried out and a capillary condensation critical point was located for a
thin Ising film \cite{28} but for a single value of $D$ only.

In the present paper, we fill in this gap by presenting a Monte Carlo study
of the critical behavior of capillary condensation in thin Ising films for a
range of thicknesses. Invoking the universality principle \cite{61}, one can
argue that nearest neighbor Ising lattices with short range surface fields
should yield the same power law, Eq.~(\ref{eq1}), as more realistic models
and real fluids in slit-like capillaries do. Unlike the situation in real
fluids, packing effects at the surfaces and a dependence of the density in
the middle of the film on its thickness are, however, absent and one focuses
on the universal critical behavior. One can argue that the order parameter
correlations in the directions parallel to the wall should all scale with
the critical exponents of the universality class of the two-dimensional
Ising model \cite{61}, 
\[
M\propto \tilde{t}^{\beta _{2}},\quad \chi \propto |\tilde{t}|^{-\gamma
_{2}},\quad \xi _{||}\propto |\tilde{t}|^{-\nu _{2}},\quad \beta
_{2}=1/8,\quad \gamma _{2}=\frac{7}{4},\quad \nu _{2}=1, 
\]
\begin{equation}
\tilde{t}\equiv 1-T/T_{c}(D)\rightarrow 0\text{,\quad all }D<\infty .
\label{eq3}
\end{equation}

Of course, one expects that for large $D$ the asymptotic critical region
where this two-dimensional critical behavior holds is very narrow, due to a
crossover to the three-dimensional critical behavior as $D\rightarrow \infty 
$, and a quantitative understanding of this crossover \cite{13,55,56,57} is
a challenging aspect of this problem, too.

In Sec.~II, we shall hence briefly define the model that is studied and the
quantities that will be analyzed and comment on the simulation methods.
Sec.~III briefly reviews the scaling predictions, including the finite-size
scaling results for the case where both $D$ and the lateral linear dimension 
$L$ are finite. Sec.~IV then presents our results on $T_{c}(D)$ and the
critical fields $H_{c}(D)$, on which our tests of Eqs.~(\ref{eq1}), (\ref
{eq2}) are based. Sec.~V discusses those aspects of our results which are
pertinent to a test of two-dimensional criticality, Eq.~(\ref{eq3}), while
Sec.~VI summarizes our conclusions.

\section{Model and simulation technique}

Invoking the well-known isomorphism between the lattice gas model of fluids
and the Ising model of magnetism (see e.g.~Ref.~\onlinecite{28} for details), 
we study
the Ising model on the simple cubic lattice in the presence of a bulk field $%
H$ and a surface field $H_{1}$, 
\begin{equation}
{\cal H}=-J\sum\limits_{\left\langle i,j\right\rangle
}S_{i}S_{j}-H\sum_{i}S_{i}-H_{1}\sum\limits_{i \, \epsilon \,{\rm surfaces}%
}S_{i},\quad S_{i}=\pm 1\text{,}  \label{eq4}
\end{equation}
where the exchange interaction $J$ is only present between nearest neighbors
on the lattice. Note that phase coexistence in the bulk (phases with
positive and negative magnetization correspond to gas and liquid phases of
the fluid, respectively) corresponds to $H=0$, see Fig.~\ref{fig1}. In the
thin film, one trivially obtains the result that for zero temperature phase
coexistence occurs for \cite{28} $H_{\text{coex}}(D,T=0)=-2H_{1}/D$, but for 
$T>0$ the variation of $H_{\text{coex}}(D,T)$ is nontrivial. While previous
work \cite{28} was mostly interested in the behavior of $H_{\text{coex}}(D,T)
$ near the temperature $T_{w}(H_{1})$ of the wetting transition \cite
{1,6,7,62}, we consider here scaled surface fields $H_{1}D^{\Delta _{1}/\nu }
$ small enough such that we stay in the nonwet regime of the surface phase
diagram of the semi-infinite system \cite{63} throughout, although we
consider the vicinity of $T_{c}^{\infty }=T_{c}(D=\infty )$. Measuring all
lengths in units of the lattice spacing, we consider film thicknesses $%
D=4,8,12,16,24,28$ and $32$ for an $L\times L\times D$ geometry, varying $L$
over an as wide range as is practical, from the point of view of available
computer ressources. In the $x,y$-directions parallel to the thin film, we
apply periodic boundary conditions as usual \cite{13,28}.

In order to be able to find $T_{c}(D)$ and $H_{c}(D)\equiv H_{\text{coex}%
}(D,T=T_{c}(D))$ reliably, we have to use aspect ratios $L/D\gg 1$. Although
the choices of film thickness as quoted above are not extremely large, it is
clear that use of rather large linear dimensions $L$ is mandatory for
obtaining reliable results. If we would use the Metropolis algorithm \cite
{64,65,66}, as done in Refs.~\onlinecite{13,28}, or the heatbath 
algorithm \cite{66}%
, ``critical slowing down'' \cite{65,66} would be a serious problem: i.e.,
the ``time'' $\tau $ over which subsequently generated system configurations
are correlated varies as \cite{66} 
\begin{equation}
\tau \propto L^{z}\text{ with }z\approx 2.16 \, (d=2)\text{ or }z\approx
2.03 \, (d=3),  \label{eq5}
\end{equation}
the prefactor in this power law being of order unity if Monte Carlo time is
measured in units of attempted Monte Carlo steps (MCS) per spin. Since we
wish to use linear dimensions of the order of $L\approx 10^{2}$, relaxation
times of the order of $10^{4}$ MCS easily result. Given the fact that
quantities like the specific heat $C_{v}$ and the susceptibility $\chi $,
recorded from fluctuations of energy and magnetization, 
\begin{equation}
C_{v}=\left( \left\langle {\cal H}^{2}\right\rangle -\left\langle {\cal H}%
\right\rangle ^{2}\right) /(L^{2}Dk_{\text{B}}T^{2}),\quad \chi =\left\{
\left\langle \left( \sum\limits_{i}S_{i}\right) ^{2}\right\rangle
-\left\langle \sum\limits_{i}S_{i}\right\rangle ^{2}\right\} /(L^{2}Dk_{%
\text{B}}T),  \label{eq6}
\end{equation}
are non-self-averaging \cite{65,66,67}, one needs $n\gg 1$ statistically
independent observations (i.e., separated by time intervals $\Delta t>\tau $%
) to obtain $C_{v}$ and $\chi $ with small enough error (the relative error
of these quantities is \cite{67} $\sqrt{2/n}$, irrespective of $L$ and $D$).
For this reason, it is clear that the use of cluster algorithms which reduce
critical slowing down \cite{66,68,69,70,71,72,73,74} is highly desirable.
However, for the present problem where both a bulk magnetic field and a
surface magnetic field of competing sign are present \{Eq.~(\ref{eq4})\}
application of cluster algorithms is nontrivial. It turns out that an
extension of the ``ghost spin algorithm'' \cite{70,71,72} to the present
problem is rather straightforwardly possible \cite{75,76}. The coupling of
spins to a magnetic field is thereby treated as if it were an additional
infinite-range exchange coupling to an extra spin $S_{G}=\pm 1$. This
coupling has the strength $h=|H|$ for spins in the interior of the film and $%
h=|H_{1}+H|$ for spins in the surface layers. In addition to putting bonds
in clusters of spins (inside a cluster all spins are connected by exchange
interactions and have the same sign) with probability \cite
{68,69,70,71,72,73,74,75,76} $p_{B}=1-\exp (-2J/k_{\text{B}}T)$ one also
puts bonds between the spins in clusters and the ghost spin $p_{G}=1-\exp
(-2h/k_{\text{B}}T)$, if the orientation of the spins in the cluster is the
same as that of the ghost spin \{which is $S_{G}=$sign $(H)$ for interior
spins and $S_{G}=$sign $(H_{1}+H)$ for spins in the surface planes,
respectively\}.

While this extension of the cluster algorithm to the case of nonzero bulk
and surface fields is formally exact, discussion of its efficiency is a
rather delicate problem: in fact, if $h/k_{\text{B}}T$ is of order unity,
also $p_{G}$ is of order unity and the infinite-range character of this
coupling then implies that huge clusters containing a large fraction of the
entire simulation volume would be created most of the time! It is clear that
under such circumstances the algorithm would be very inefficient; as in the
case of zero field it is necessary for a good performance of a cluster
algorithm that typically large clusters are created but a single large
cluster must contain only a negligible fraction of the total volume in the
thermodynamic limit. As a consequence, one needs $h/k_{\text{B}}T\ll 1$, and
%
since $k_{\text{B}}T$ is in the range of $3.5 - 4.5$ we have thus
chosen to work with a single value of the surface field, namely 
$H_{1}=-0.015J$. Even for this small value -- note that the corresponding
value of $H$ is typically one or two orders of magnitude smaller, see below
-- the performance of the algorithm has significantly deteriorated, in
comparison with the case without any magnetic fields. This fact can be
clearly demonstrated by a binning analysis \cite{74,77} of the magnetization 
$m$ in the system: the $N$ (dynamically correlated) subsequent observations $%
m_{\nu }=(1/N)\sum\limits_{i=1}^{N}S_{i}^{\nu }$ are grouped into $n=N/N_b$
blocks of length $N_b$, from which block averages $\tilde{m}_{\mu }$ of the
corresponding $N_b$ observations \{$m_{\nu }$\} belonging to the block with
index $\mu $ are formed. Then 
\begin{equation}
\Delta m\equiv \lbrack n(n-1)]^{-1/2}\sqrt{\sum\limits_{i=1}^{n}(\tilde{m}%
_{\mu }-\bar{m})^{2}}  \label{eq7}
\end{equation}
(where $\bar{m}=n^{-1}\sum\limits_{i=1}^{n}\tilde{m}_{i})$ is studied as a
function of block length $N_b$ (Fig.~\ref{fig2}): when $\Delta m$ is
independent of $N_b$, the subsequent $\tilde{m}_{\mu }$ are statistically
independent, and $\Delta m$ is a good estimation of the statistical error;
otherwise one sees a systematic increase of $\Delta m$ with $N_b$ and the
value of $N_b$ needed to reach a saturation value yields an estimate for the
correlation time. For the Metropolis algorithm and the chosen system size ($%
L $=128), even for $N_b=4000$ one is far from saturation, and hence it is
clear that this algorithm would be very impractical for the present problem.
For the cluster algorithm and $H_{1}=0$, on the other hand, $\Delta m$ vs. $N_b
$ is essentially constant, $\Delta m\approx 0.0004$, the correlation time
being of order unity, as expected \cite{66,68,73,74}. However, this is not
so for the cluster algorithm in the case $H_{1}=-0.015$: $\Delta m$
saturates at a plateau of about $\Delta m\approx 0.007$, i.e. the error is
almost a factor 20 larger, and the correlation time is of the order of $\tau
_{m}\approx 280$ Monte Carlo steps in the example shown in Fig.~\ref{fig2}%
(b). Thus, while the gain of the cluster algorithm in the zero field case
compared to the Metropolis algorithm is very significant, in our problem it
is only rather modest! This - somewhat unexpected - dramatic decrease of the
efficiency of the cluster algorithm with increasing strength of the surface
(and bulk) fields has prevented us both from studying systems larger than $%
L=128$ and from studying the dependence on $H_{1}$ systematically. Runs of
length up to 1.2 million Monte Carlo steps (MCS) were performed.

As is well known \cite{66,74,78}, cluster algorithms at critical points of
Ising systems are rather sensitive to correlations among the pseudorandom
numbers produced by the random number generator. In the present work, we
used thus an improved version of the standard ''R250'' generator \cite{79},
where two versions \{one based on the pair of integers (250, 103) the other
with the pair (521, 168)\} are combined with the logical XOR operation.

In order to make best use of our simulation data, we apply standard
multihistogram interpolation techniques \cite{66,73,74,80}. Note that we
needed a three-dimensional histogram $P(E,m,m_{1})$, $E$ being the exchange
energy, $m_{1}$ the magnetization in the surface plane, in order to allow
reweightings in the full parameter space of independent control variables ($%
T,H,H_{1}$) and hence the storage requirements for $P$ are nontrivial.
However, noting that all measurements of $E,m,m_{1}$ can be represented by
integers, each integer needing 4 Byte, we can store the time series of 10$%
^{6}$ observations with a storage of 12 MByte, irrespective of the choices
of $L$ and $D$.

The multihistogram reweighting with respect to the bulk field $H$ is crucial
in order to be able to find the field $H_{\text{coex}}(T)$, along which for $%
T<T_{c}(D)$ two-phase coexistence occurs, applying the ``equal weight rule'' 
\cite{65,66,74,81}: In the space of variables $(E,m,m_{1})$, the two phases
show up as separate peaks of $P(E,m,m_{1})$ \{or $P(m,m_{1})$, respectively,
see Fig.~\ref{fig3}(a), when one studies an isotherm one can integrate out $%
E $, of course\}, which have precisely the same weight at $H=H_{\text{coex}%
}(T) $ while for $H\neq H_{\text{coex}}(T)$ (but not too far away from it)
the two peaks can still be identified but have different weight. With the
multihistogram reweighting, a small number of simulation points suffices to
generate the curve $H_{\text{coex}}(T)$ \{and its extrapolation into the
regime $T>T_{c}(D)$\} with reasonable precision, see Fig.~\ref{fig3}(b).
Since near $T_{c}(D)$ the free energy differences between the two phases are
very small also off coexistence, the statistical error in the estimation of $%
H_{\text{coex}}(T)$ is not negligible, and also systematic errors, since $%
L/D $ is not large enough, need to be considered. The latter problem also
affects the estimation of $T_{c}(D)$, as will be discussed in Sec. IV.

\section{SCALING PREDICTIONS}

For completeness, we first summarize the pertinent predictions of the
scaling theory for thin Ising films near the critical point \cite
{16,17,18,19,83}, assuming the lateral linear dimension $L$
infinite, and consider the extension \cite{57} to finite $L$ in the
following. The singular part of the free energy per spin is assumed to scale
as follows 
\begin{equation}
f_{\text{sing}}\left( D,T,H,H_{1}\right) \approx |t|^{2-\alpha }\widetilde{f}%
_{\pm }\left( D|t|^{\nu }~,~H|t|^{-\triangle },~H_{1}|t|^{-\triangle
_{1}}\right) ~,  \label{eq8}
\end{equation}
where $\alpha $ is the exponent of the specific heat of the 
three-dimensional Ising model, $t=\left( T-T_{c}\left( \infty \right) \right)
/T_{c}\left( \infty \right) ,~\widetilde{f}_{\pm }$ is a ``scaling
function'' (with two different branches, referring to the sign of $t$), and
the other exponents have already been defined in Sec. I.

Now it is convenient to introduce the scaling variables

\begin{equation}
x\equiv D|t|^{\nu }~,~w\equiv H_{1}D^{\Delta _{1}/\nu },  \label{eq9}
\end{equation}
and then Eq.~(\ref{fig8}) can also be written as, eliminating $|t|$ from the
arguments of $\widetilde{f}_{\pm }$,

\begin{equation}
f_{\text{sing}}\left( D,T,H,H_{1}\right) \approx |t|^{2-\alpha }\widetilde{f}%
_{\pm }\left( x,\frac{HD^{\Delta /\nu }}{x^{\Delta /\nu }}~,\frac{w}{%
x^{\Delta _{1}/\nu }}\right) .  \label{fig10}
\end{equation}

Since the critical point of the thin film is shifted relative to the bulk
critical point $T_{c}\left( \infty \right) $, it must correspond to a
singular behavior of the scaling function $\widetilde{f}_{\pm }$. At fixed $%
H_{1}$ and fixed $D$ this means the scaling function $\widetilde{f}_{\pm }$ $%
\left( x,y,y_{1}=w/x^{\Delta _{1}/\nu }\right) $ has a singularity at a
point $x_{c}\left( w\right) ,$ $y_{c}\left( w\right) $. Therefore the shifts 
$\Delta T_{c}\left( D\right) ,\Delta H_{c}\left( D\right) $ follow as \cite
{16} 
\begin{equation}
\Delta T_{c}=T_{c}\left( D,H_{1}\right) -T_{c}\left( \infty \right)
=-B_{T}D^{-1/\nu }X_{c}\left( CH_{1}D^{\Delta _{1}/\nu }\right) \quad ,
\label{eq11}
\end{equation}
\begin{equation}
\Delta H_{c}\equiv H_{c}(D,H_{1})=-B_{H}D^{-\Delta /\nu
}Y_{c}(CH_{1}D^{\Delta _{1}/\nu }).  \label{eq12}
\end{equation}
The scaling functions $X_{c},Y_{c}$ are universal, while $B_{T},B_{H}$ and $%
C $ are non-universal critical amplitudes, which are normalized such that $%
X_{c}(Cw)=1+(Cw)^{2}+\ldots ,$ $Y_{c}(Cw)\approx Cw+o(Cw)^{3}$. Note that
both functions are analytic for $w\rightarrow 0$, and $\Delta T_{c}$ should
be an even function of $H_{1}$ and hence $w$, while $\Delta H_{c}$ must be
an odd function of $H_{1}$. From these considerations, for small $H_{1}$
Eqs.~(\ref{eq1}), (\ref{eq2}) result, remembering \cite{28} that $\mu
_{c}(D)-\mu _{c}(\infty )=-2H_{c}$.

An alternative argument for Eq.~(\ref{eq1}), which also elucidates how this
equation fits together with the Kelvin equation $(H_{c}\propto -H_{1}/D)$ in
the critical region for large enough $D$, derives from a consideration of
phase coexistence for temperatures slightly below $T_{c}(D)$. If $H_{1}=H=0$%
, we would have two coexisting phases with magnetization profiles $m^{+}(z)$
and $m^{-}(z)=-m^{+}(z)$ across the film, and both states have the same free
energy $F_{+}(0,0)=F_{-}(0,0)$. Since these profiles are smooth functions of 
$H$ and $H_{1}$, an expansion of the free energies around $F_{+}(0,0)$ \{or $%
F_{-}(0,0)$, respectively\} yields 
\begin{equation}
-\Delta F_{+}\equiv F_{+}(0,0)-F_{+}(H,H_{1})=\overline{m^{+}}%
HDL^{2}+2m_{1}^{+}H_{1}L^{2},  \label{eq13}
\end{equation}
\begin{equation}
-\Delta F_{-}\equiv F_{-}(0,0)-F_{-}(H,H_{1})=\overline{m^{-}}%
HDL^{2}+2m_{1}^{-}H_{1}L^{2},  \label{eq14}
\end{equation}
where $\overline{m^{+}},\overline{m^{-}}$ refer to the average over the
magnetization profile in the respective states, and $m_{1}^{+},m_{1}^{-}$
the layer magnetizations in the surface layer. To leading order for small $%
H_{1}$ and small $H$ in Eqs.~(\ref{eq13},\ref{eq14}), $\overline{m^{+}},%
\overline{m^{-}}$ and $m_{1}^{+},m_{1}^{-}$ are to be taken at zero fields,
and thus satisfy the symmetry $\overline{m^{-}}=-\overline{m^{+}}$ and $%
m_{1}^{-}=-m_{1}^{+}$. Phase coexistence occurs for $\Delta F^{+}=\Delta
F^{-}$, and hence the Kelvin equation results, 
\begin{equation}
H_{\text{coex}}(D,T,H_{1})\approx -\frac{2H_{1}}{D}\frac{m_{1}^{+}(D,T)}{%
\overline{m^{+}}(D,T)}.  \label{eq15}
\end{equation}
Assuming then that $D\gg \xi $, the correlation length in the bulk, $m^{+}(z)
$ will approach the bulk spontaneous magnetization $m_{b}=\hat{B}%
_{b}(-t)^{\beta }$ almost everywhere, and hence $\overline{m_{+}}\approx 
\hat{B}_{b}(-t)^{\beta }$, $\hat{B}_{b}$ being the respective critical
amplitude. Likewise $m_{1}^{+}(D,T)$ approaches the surface layer
magnetization of a semi-infinite system \cite{19,50,83} $m_{1}=\hat{B}%
_{1}(-t)^{\beta _{1}}$, with $\hat{B}_{1}$ the corresponding amplitude.
Therefore Eq.~(\ref{eq15}) becomes in this limit 
\begin{equation}
H_{\text{coex}}(D,T,H_{1})\approx -\frac{2H_{1}}{D}\frac{\hat{B}_{1}}{\hat{B}%
_{b}}(-t)^{\beta _{1}-\beta },\quad x=D|t|^{\nu }\rightarrow \infty .
\label{eq16}
\end{equation}
Since \cite{49} $\beta \approx 0.325$ and \cite{19,50,51,52} $\beta
_{1}\approx 0.78-0.80$, we see that the coefficient of the term $H_{1}/D$ in
Eq.~(\ref{eq16}) gets smaller and smaller as $|t|$ gets smaller. Due to this
vanishing coefficient in the limit $|t|\rightarrow 0$ a smooth crossover
between the Kelvin equation, Eqs.~(\ref{eq15}, \ref{eq16}), and Eq.~(\ref
{eq1}) becomes possible. Remembering that for $D$ finite there is a shift of 
$T_{c}$ as given by Eq.~(\ref{eq2}), we can further conclude that the
critical field should be of the order 
\begin{equation}
H_{c}(D_{1},H_{1})\equiv H_{\text{coex}}(D,T_{c}(D),H_{1})\propto -\frac{%
2H_{1}}{D}D^{-(\beta _{1}-\beta )/\nu }=-2H_{1}D^{-(\Delta -\Delta _{1})/\nu
}\quad ,  \label{eq17}
\end{equation}
where in the last step the standard scaling relations $\beta =2-\alpha
-\Delta ,\beta _{1}=2-\alpha -\nu -\Delta _{1}$ were used. Eq.~(\ref{eq17})
obviously is nicely compatible with Eq.~(\ref{eq1}).

Eq.~(\ref{eq16}) allows an interesting conclusion to be drawn on the slope
of the coexistence curve at the critical point of the film (cf.~Fig.~\ref
{fig1}). We find first for the angle $\theta (t)$ describing this slope for $%
T<T_{c}(D)$%
\begin{equation}
\tan (\theta )\equiv (\partial H/\partial T)_{H_{1}}=\frac{2H_{1}}{%
T_{c}(\infty )}\frac{\beta _{1}-\beta }{D}\frac{\hat{B}_{1}}{\hat{B}_{b}}%
(-t)^{\beta _{1}-\beta -1},x=D|t|^{\nu }\rightarrow \infty ~.  \label{eq18}
\end{equation}
Since $\beta _{1}<1+\beta $, the exponent of $(-t)$ is negative, and thus
for the considered limit the slope diverges (i.e., varying $t$ at very large
but fixed $D$). However, this limit does not include the limiting slope at $%
T\rightarrow T_{c}(D)$ itself, since then Eqs.~(\ref{eq2}), (\ref{eq11}) yield 
$t\propto D^{-1/\nu }$, and hence 
\begin{equation}
\tan (\theta )\propto H_{1}D^{-(\beta _{1}-\beta -1)/\nu
-1}=H_{1}D^{-(\Delta -\Delta _{1}-1)/\nu }.  \label{eq19}
\end{equation}
In Landau theory, $\Delta =3/2,\Delta _{1}=1/2$ and hence the power of $D$
vanishes, i.e., the slope is nonzero and finite at $T_{c}(D)$ in the limit $%
D\rightarrow \infty $. For the three-dimensional Ising model, the best
available exponent estimates \cite{49,19,50,51,52} imply $(\Delta -\Delta
_{1}-1)/\nu \approx 0.12-0.16$, i.e. $\theta \rightarrow 0$ for $%
D\rightarrow \infty $! This result also implies that for capillary
condensation field mixing effects \cite{84} are asymptotically not very
important.

We now briefly consider the extension of the scaling theory to include
finite-size effects due to the finite lateral linear dimension $L$ \{in
Eqs.~(\ref{eq13}, \ref{eq14}) we have assumed the limit $L\rightarrow \infty 
$ throughout\}. This can simply be done by including the aspect ratio $L/D$
as an additional scaling variable in Eqs.~(\ref{eq8}, \ref{eq9}), which we
then rewrite as follows 
\begin{equation}
f_{\text{sing}}(D,T,H,H_{1},L)\approx D^{-3}\tilde{f}(D^{1/\nu
}t,L/D,HD^{\Delta /\nu },H_{1}D^{\Delta _{1}/\nu }).  \label{eq20}
\end{equation}
Since for finite $L$ the free energy and its derivatives are smooth
functions of $t$, it is more convenient to use $D^{1/\nu }t$ rather than $%
x=D|t|^{\nu }$ as a scaling variable. From Eq.~(\ref{eq20}), we immediately
obtain the following scaling results for the specific heat, the
magnetization and the susceptibility of the thin film 
\begin{equation}
C_{v}=D^{\alpha /\nu }\tilde{C}(D^{1/\nu }t,L/D,HD^{\Delta /\nu
},H_{1}D^{\Delta _{1}/\nu }),  \label{eq21}
\end{equation}
\begin{equation}
m=D^{-\beta /\nu }\tilde{m}(D^{1/\nu }t,L/D,HD^{\Delta /\nu },H_{1}D^{\Delta
_{1}/\nu }),  \label{eq22}
\end{equation}
\begin{equation}
\chi =D^{\gamma /\nu }\tilde{\chi}(D^{1/\nu }t,L/D,HD^{\Delta /\nu
},H_{1}D^{\Delta _{1}/\nu }),  \label{eq23}
\end{equation}
where $\tilde{C},\tilde{m}$ and $\tilde{\chi}$ are appropriate scaling
functions. Since we choose $H_{1}$ fixed, $D$ fixed, $H_{1}D^{^{\Delta
_{1}/\nu }}\ll 1$, and $H$ is chosen according to Eqs.~(\ref{eq16}, \ref
{eq17}) \{in practice this is done by applying the reweighting technique and
the equal area rule, cf. Fig.~\ref{fig3}\} the last two arguments $%
HD^{\Delta /\nu },H_{1}D^{\Delta _{1}/\nu }$ in Eqs.~(\ref{eq21})--(\ref{eq23}%
) can be ignored in the following discussion.

Now in the limit $L\rightarrow \infty $ we expect that $C_{v}$ exhibits a
logarithmic singularity for $T\rightarrow T_{c}(D)$, while the critical part
of the magnetization $m_{\text{crit}}\equiv m-m(T_{c}(D),H,H_{1})$ should
behave as \cite{61} 
\begin{equation}
m_{\text{crit}}\propto \lbrack 1-T/T_{c}(D)]^{\beta _{2}},\quad \beta
_{2}=1/8,  \label{eq24}
\end{equation}
and the susceptibility 
\begin{equation}
\chi \propto |1-T/T_{c}(D)|^{-\gamma _{2}},\quad \gamma _{2}=7/4.
\label{eq25}
\end{equation}
For finite $L$, however, these singularities are all rounded off and we
rather expect that both $C_{v}$ and $\chi $ exhibit maxima of finite height
at temperatures $T_{\max }^{c}$ $\left( D\right) $, $T_{\max }^{\chi }\left(
D\right) $. From Eqs.~(\ref{eq21}), (\ref{eq23}), we readily predict (in the
limit $|H_{1}|D^{\Delta _{1}/\nu }\ll 1$) 
\begin{equation}
\frac{T_{\max }^{c}\left( D\right) -T_{c}(D)}{T_{c}(\infty )}D^{1/\nu
}=\Delta \widetilde{T}_{\max }^{c}\left( L/D\right) ,  \label{eq26}
\end{equation}
\begin{equation}
\frac{T_{\max }^{\chi }\left( D\right) -T_{c}(D)}{T_{c}(\infty )}D^{1/\nu
}=\Delta \widetilde{T}_{\max }^{\chi }\left( L/D\right) ,  \label{eq27}
\end{equation}
with $\Delta \widetilde{T}_{\max }^{c}\left( L/D\right) $, $\Delta 
\widetilde{T}_{\max }^{\chi }\left( L/D\right) $ being suitable scaling
functions that describe the shift of these maxima positions as functions of
the aspect ratio $L/D$. From this analysis one also can predict \cite{57}
how the height of the maxima should depend on $D$ and $L$, for $L>>D$

\begin{equation}
C_{v}^{\max }\propto D^{\alpha /\nu }\ln (L/D),  \label{eq28}
\end{equation}
\begin{equation}
\chi ^{\max }\propto D^{\gamma /\nu -7/4}\text{ \ }L^{7/4}\text{,}
\label{eq29}
\end{equation}
and how the absolute value of the order paramter should decrease at $%
T_{c}(D) $, 
\begin{equation}
\left\langle |m_{\text{crit}}|\right\rangle _{T=T_{c}\left( D\right)
}\propto D^{1/8-\beta /\gamma }\quad L^{-1/8}\quad .  \label{eq30}
\end{equation}
Finally, in the limit $L\rightarrow \infty $ the $D$-dependence of the
critical amplitudes associated with the two-dimensional critical behavior
%
(see Ref.~\onlinecite{57} for a more detailed discussion), defining now $%
\widetilde{t}=\left[ T-T_{c}\left( D\right) \right] /T_{c}\left( \infty
\right) $, can be read off from the following equations,

\begin{equation}
C_{v}\propto D^{\alpha /\nu }\ln |\widetilde{t}|\quad ,  \label{eq31}
\end{equation}
\begin{equation}
m_{\text{crit}}\propto D^{\left( 1/8-\beta \right) /\nu }\left( -\widetilde{t%
}\right) ^{1/8}\quad ,  \label{eq32}
\end{equation}
\begin{equation}
\chi \propto D^{\left( \gamma -7/4\right) /\nu }|\widetilde{t}|^{-7/4}\text{
\ }.  \label{eq33}
\end{equation}
Due to the crossover scaling between two- and three-dimensional critical
behavior, that Eqs.~(\ref{eq20}), (\ref{eq21}), (\ref{eq22}), and (\ref{eq23})
describe, a singular dependence of the various critical amplitudes on film
thickness results at the capillary condensation critical point.

\section{Numerical Results on $T_{c}\left( D\right) $ and $H_{c}\left(
D\right) .$}

For locating critical points in the bulk, a convenient method is to study
the fourth order cumulant of the order parameter

\begin{equation}
U_{L}\left( T\right) =1-\left\langle m_{\text{crit}}^{4}\right\rangle /\left[
3\left\langle m_{\text{crit}}^{2}\right\rangle ^{2}\right]  \label{eq34}
\end{equation}
for a range of linear dimensions $L$ as a function of temperature, and to
look for a common intersection point \cite{85} \{which for the universality\
class of the two-dimensional Ising model, should have the value \cite{86} $%
U^{\ast }=U_{L}\left( T=T_{c}\right) $ $=0.610690\left( 1\right) $\}. In our
case, we have to follow a path along $H=H_{\text{coex}}\left( T\right) $
\{as shown in Fig.~\ref{fig3}(b)\} when we record these cumulants, and since
this path is not exactly known but only within some numerical error, it is
clear that this method is more difficult to apply than for ordinary bulk
Ising models. In addition, even for small $D$ the data are plagued by
crossover scaling effects (Fig.~\ref{fig4}): the curves for the values of $L$
that are practically available do not intersect in a common point, but
rather the intersection points are scattered and fall below the theoretical
value $U^{\ast }$. This failure of verifying the common intersection points
is not unexpected, since Fig.~\ref{fig4} includes data for which the aspect
ratio $L/D$ is as small as 4 (a) or even 2 (b), rather than only data for
which $L/D\gg 1$. In fact, from the treatment of the previous section we can
readily conclude that 
\begin{equation}
U_{L}\left( T=T_{c}\left( D\right) \right) =\widetilde{U}\left( L/D\right)
\label{eq35}
\end{equation}
and only in the limit $L/D\rightarrow \infty $ shall we have $U\left( \infty
\right) =U^{\ast }$.

An alternative and widely used recipe to find the critical temperature is to
try an extrapolation of the maxima of the specific heat and susceptibility
versus $L^{-1/\nu }$or of the cumulant intersection points. Considering the
intersection of $U_{L}\left( T\right) $ and $U_{bL}\left( t\right) $ with a
scale factor $b>1$, it can be argued \cite{85} that corrections to 
finite-size scaling lead to a shift of the intersection point that varies 
with $b$
proportional to $\left[ b^{1/\nu }-1\right] ^{-1}$ for large $b$. Fig.~\ref
{fig5} shows some attempts to carry out such extrapolations, again for $D=8$
%
and $D=28$ (data for all other choices of $D$ can be found in \cite{75} and
look similar). These figures show that $T_{\max }^{c}$ approaches $%
T_{c}\left( D\right) $ in a non-monotonic fashion, and also the curve $%
T_{\max }^{\chi }$ vs.\ $L^{-1/\nu }$ is distinctly non-linear. Fitting
asymptotic straight lines to both data sets one obtains results for $T_{c}\left(
D\right) $ that are roughly compatible with each other, and with the
(linear) extrapolation of the cumulant intersections. Although the accuracy
of $T_{c}\left( D\right) $ obtained in this way is several orders\ of
magnitude less than in the case of the bulk three-dimensional Ising model 
\cite{87}, the data are accurate enough to allow a useful test of Eqs.~(\ref
{eq1}), (\ref{eq2}).

The consistency of our analysis can be checked further by testing for the
scaling behavior predicted in Eqs.~(\ref{eq26}), (\ref{eq27}), see Fig.~\ref
{fig6}. Here all data points are included for all values of $D$ and $L$ that
have been studied and $T_c(D)$ is chosen such that the best data collapse 
is achieved. It is seen that the non-monotonic variation of the
temperature at which the specific heat has its maximum is an intrinsic
property of this scaling function describing the system shape effects in
terms of the aspect ratio $D/L$ of the simulation box. The interpolating curves
are simple parabolic fits which translate back into the solid lines in the
left part of Fig.~\ref{fig5}.

The values of $T_{c}\left( D\right) $ that we have determined as shown in
Figs.~\ref{fig5}, \ref{fig6} are collected in Table I, which includes also
our estimates for $H_{c}\left( D\right) $. Log-log plots of these data
versus $D$ almost look like straight lines, however, there is a slight but
systematic curvature, and if this curvature were disregarded and straight
lines were fitted to all the data nevertheless, the resulting effective
exponents would systematically deviate from the theoretial predictions in
Eqs.~(\ref{eq1}), (\ref{eq2}).

Better results are obtained if one fits effective exponents from successive
thicknesses only $\left( D=4,8,12;D=8,12,16;\ldots ;D=24,28,32\right) $,
which can be extrapolated vs.\ $1/D$ reasonably well, and converge nicely
towards the theoretical predictions (Fig.~\ref{fig7}), namely $-1/\nu
\approx -1.587$ and $-\left( \Delta -\Delta _{1}\right) /\nu \approx -1.75$.
Conversely, if Fig.~\ref{fig7}(b) was taken as an independent estimation of
the exponent $\Delta _{1}$, we would obtain $\Delta _{1}=0.459(13)$, which
indeed is compatible with the existing recent estimates \cite{50,51,52}.

\section{A Test of Two-Dimensional Critical Behavior}

In this section, we are concerned with the question to what extent the data
provide some evidence for the prediction (Eq.~\ref{eq3}) that the capillary
condensation critical point displays critical exponents of the
two-dimensional Ising universality class. Since the accessible values of the
lateral linear dimension $L$ are not very large, however, we cannot expect
that a regime can be reached where the parallel correlation length $\xi
_{||} $ satisfies the criterion $D\ll \xi _{||}\ll L$ - only in such a
regime a direct observation of these power laws would be possible. Hence we
attempt to study the critical behavior again via a finite-size scaling
analysis, using \cite{10,11,12,13,14,18,53,54,65,66},

\begin{equation}
\left\langle |m_{\text{crit}}|\right\rangle L^{v}=\widetilde{M}\left( L^{u}%
\widetilde{t}\right) \quad,  \label{eq36}
\end{equation}
\begin{equation}
\chi L^{-w}=\widetilde{\chi }\left( L^{u}\widetilde{t}\right)\quad ,  \label{eq37}
\end{equation}
where the exponents $u,v,w$ should take the values 
\begin{equation}
u=1/\nu _{2}=1,~v=\beta /\nu _{2}=1/8,~w=\gamma _{2}/\nu _{2}=7/4.
\label{eq38}
\end{equation}
Eqs.~(\ref{eq36}), (\ref{eq37}), and (\ref{eq38}) are appropriate if $D\ll \xi
_{||}$ still holds but $\xi _{||}$ and $L$ are of the same order. In
practice, however, also the condition $D\ll \xi _{||}$ is hard to satisfy
since we wish to include some data for which $L/D$ is not very large. It
then helps to relax the theoretical condition, Eq.~(\ref{eq38}), and rather
treat $u,v,w$ as effective exponents \cite{57}: in this way, one can take
into account to some extent the corrections to finite-size scaling arising
from the crossover between two- and three-dimensional Ising critical
behavior.

Fig.~\ref{fig8} shows that this procedure works reasonably well, and Table
II gives a listing of the fit exponents $u,v,w$, and corresponding effective
exponents $\nu _{\text{eff}}=1/u$, $\beta _{\text{eff}}=v/u$ and $\gamma _{%
\text{eff}}=w/u$. It is seen from Table II that both $u,v$ and $w$ gradually
change from the two-dimensional values towards the three-dimensional ones,
although even for $D=32$ one is still far away from the theoretical values
for the latter. While $\beta _{\text{eff}}$ has increased significantly, $%
\gamma _{\text{eff}}$ within the accuracy of this estimation has hardly
changed at all. If we consider an effective dimensionality from the
hyperscaling relation \cite{61}, defined as $d_{\text{eff}}=(\gamma _{\text{%
eff}}+2\beta _{\text{eff}})/\gamma _{\text{eff}}=w+2v$, we find $d_{\text{eff%
}}=2.0\pm 0.15$, and there is no systematic trend with $D$. While the latter
observation is in accord with a previous study using ``neutral walls'' \cite
{57}, where $H_{c}(D)\equiv 0$, we have obtained in the present work a much
better evidence that for small $D$ the behavior is compatible with
two-dimensional Ising criticality than was possible in the latter model \cite
{57}. Note also that in the present study there is a rather broad range of $%
D $ where $\nu _{\text{eff}}>1$, which was not the case in \cite{57}. Due to
the systematic problems of fitting several effective exponents from somewhat
noisy data and the restricted range over which this fit is applicable we do
not think that these discrepancies are a proof of non-universal crossover
behavior, however. We feel that this problem needs a more careful study.

\section{CONCLUSIONS}

In this paper Monte Carlo simulations have been presented attempting to test
theoretical predictions about the capillary condensation critical point.
Using an extension of the Swendsen-Wang cluster algorithm including
competing surface and bulk magnetic fields, for Ising films with thicknesses 
$D=4,8,12,16,24,28$ and $32$ the critical temperature $T_{c}(D)$ and
corresponding critical field $H_{c}(D)$ for a surface magnetic field $H_{1}$
have been estimated. The data are compatible with the power laws presented
about 20 years ago by Fisher and Nakanishi. Also the expected
two-dimensional critical behavior is compatible with our data, though the
accuracy of the resulting effective exponents is rather low (Table II) and
hence a more convincing proof would be desirable, but is not feasible with
the present computer ressources.

A challenging problem that we have not solved is the development of an
efficient version of the cluster algorithm that allows to work with surface
and bulk fields that are not extremely small. The algorithm that we have
used was much less efficient even for $H_{1}=-0.015J$ than for $H_{1}=0$,
and a study of capillary condensation critical points over the range where $%
(H_{1}/J)D^{\Delta _{1}/\nu }$ is not small, and hence the nonlinear part of
the scaling function $Y_{c}(CH_{1}D^{\Delta _{1}/\nu })$ would be probed,
turned out not to be feasible either. Thus, in spite of a longstanding
effort to deal with theory and simulation of capillary condensation there
remain still some missing links. A particularly intriguing problem is to
elucidate the crossover between three-dimensional and two-dimensional
critical behavior in these thin films. Finally, it is hoped that the present
study provides an incentive to address this problem also by suitable
experiments.

\underline{Acknowledgements}: One of us (O. D.) acknowledges support from
the Deutsche\ Forschungsgemeinschaft (DFG) under grant No. Bi314/16, another
(W. J.) acknowledges support from the Heisenberg program of the DFG and from
Sonderforschungsbereich 262/D1.

\newpage

TABLE I: \underline{Critical temperatures and fields}

\begin{tabular}{ccc}
$D$ & $T_{c}(D)$ & $H_{c}(D)/J$ \\ \hline
4 & 3.8705(3) & 0.006644(32) \\
8 & 4.2409(3) & 0.002528(14) \\
12 & 4.3561(3) & 0.001367(10) \\
16 & 4.4084(3) & 0.000867(6) \\
24 & 4.4549(5) & 0.000448(3) \\
28 & 4.4665(4) & 0.000348(3) \\
32 & 4.4749(5) & 0.000279(3)
\end{tabular}

\begin{tabular}{ccc}
$\infty $ & 4.51152(2) \ \ \ \ \  & \ \ \ \ \ \ \ \ 0
\end{tabular}

\bigskip

TABLE II: \underline{Effective exponents for order parameter and
susceptibility}

\bigskip
\begin{tabular}{ccccccc}
$D$ & $u$ & $v$ & $w$ & $\beta _{\text{eff}}$ & $\gamma _{\text{eff}}$ & $\nu _{
\text{eff}}$ \\
2-dim & 1 & 1/8 & 7/4 & 1/8 & 7/4 & 1 \\ \hline
4 & 0.956 & 0.126 & 1.67 & 0.132 & 1.75 & 1.064 \\
8 & 1.018 & 0.136 & 1.72 & 0.133 & 1.69 & 0.982 \\
12 & 0.944 & 0.138 & 1.67 & 0.146 & 1.77 & 1.059 \\
16 & 0.938 & 0.139 & 1.61 & 0.148 & 1.72 & 1.066 \\
24 & 0.898 & 0.145 & 1.53 & 0.161 & 1.70 & 1.114 \\
28 & 0.853 & 0.141 & 1.48 & 0.165 & 1.74 & 1.172 \\
32 & 0.884 & 0.155 & 1.54 & 0.175 & 1.74 & 1.131 \\ \hline
3-dim & 1.587 & 0.518 & 1.96 & 0.327 & 1.24 & 0.630
\end{tabular}

\begin{figure}[tbp]
\caption{Schematic phase boundary for an Ising film of thickness $D$, where
on both surfaces a field $H_{1\text{ }}$acts, in the plane of variables
temperature $T$ and bulk field $H$.}
\label{fig1}
\end{figure}
\begin{figure}[tbp]
\caption{Error $\Delta m$ as calculated from Eq.~(\ref{eq7}) plotted vs.
block length $N_b$ for the case $D=32$, $L=128$, and two choices of $%
H_{1},H_{1}=0$ (a) and $H_{1}=-0.015J$ (b). In both cases the chosen
temperature (and bulk field $H$ in the case of (b)) are adjusted such that
the system is precisely at the critical point. Upper curve in each panel
represents the Metropolis algorithm, lower curve represents the cluster
algorithm.}
\label{fig2}
\end{figure}
\begin{figure}[tbp]
\caption{(a) Unnormalized histogram $P\left( m,m_{1}\right) $ of the system
with $D=32$, $L=96,$ $H_{1}/J=-0.015,H/J=0.00028$ at $k_{B}T/J=4.471$, which
is a state close to the two-phase coexistence line. (b) Two-phase
coexistence line in the plane of variables $H/J$ and $k_{\text{B}}T/J$ for $%
D=28$, estimated separately for four different choices of $L$ from the
``equal weight''-rule, showing also the statistical errors as estimated from
Jackknife procedures \protect\cite{82}. The two vertical lines show the
error interval of the critical temperature.}
\label{fig3}
\end{figure}
\begin{figure}[tbp]
\caption{Cumulants $U_{L}(T)$ plotted vs. $T$ for $D=8$ (a) and $D=28$ (b),
for various choices of $L$ as indicated in the figures. Dotted horizontal
straight lines indicate the theoretical value $U^{\ast }$ taken from
\protect\cite{86}.}
\label{fig4}
\end{figure}
\begin{figure}[tbp]
\caption{Temperatures of specific heat and susceptibility maxima (left part)
plotted vs. $L^{-1/\protect\nu }$, and temperatures of cumulant
intersections plotted vs. $(b^{1/\protect\nu }-1)^{-1}$ (right part), for $%
D=8$ (a) and $D=28$ (b). In the left part the dashed curves show straight line 
fits and the solid curves correspond to the master curves in Fig.~\ref{fig6}.}
\label{fig5}
\end{figure}
\begin{figure}[tbp]
\caption{Master curves for temperature of the susceptibility maxima (upper
part) and specific heat maxima (lower part) plotted vs. the inverse aspect
ratio.}
\label{fig6}
\end{figure}
\begin{figure}[tbp]
\caption{Plot of $-1/\protect\nu_{\text{eff}}$ (a) and $-\left[ (\Delta
-\Delta _{1})/\protect\nu \right]_{\text{eff}}$ (b) vs. $1/D$ (effective
exponents were fitted from three successive values of $D$, cf. text).}
\label{fig7}
\end{figure}
\begin{figure}[tbp]
\caption{(a) Finite-size scaling plot for the critical part of the
magnetization, $\left\langle |m_{crit}|\right\rangle $, for $D=4$ and four
choices of $L$ as indicated, using $T_{c}(D)$ as quoted in Table I, and
effective exponents $u=0.956$, $v=0.126$. The straight line has a slope
indicating the exponent $\protect\beta _{2}=1/8$. (b) Same as (a) but for $%
D=32$, using now $u=0.884$, $v=0.155$. (c) Same as (b) but for the
susceptibility, using $u=0.884$, $w=1.55$. The straight line indicates the
exponent $\protect\gamma _{2}=7/4$.}
\label{fig8}
\end{figure}

\clearpage
\newpage

\setlength{\textheight}     {297mm}
\setlength{\textwidth}      {200mm}
\setlength{\topmargin}      {0mm}
\setlength{\evensidemargin} {0mm}
\setlength{\oddsidemargin}  {0mm}
\setlength{\footskip}       {0mm}
\setlength{\unitlength}     {1mm}
\setlength{\parindent}{0pt}
\setlength{\parskip}{5pt}
\setlength{\voffset} {-1.0in}

\renewcommand{\topfraction}{1.0}
\renewcommand{\bottomfraction}{1.0}
\renewcommand{\textfraction}{0}
\renewcommand{\baselinestretch}{1.2}

\frenchspacing


\input{epsf}

\def\figurename{\footnotesize{\bf Abb.}}
\def\tablename {\footnotesize{\bf Tab.}}


\sloppy


\thispagestyle{empty}
\vspace*{3cm}
\begin{figure} [hbt]
  \begin{center}
    \epsfig{file=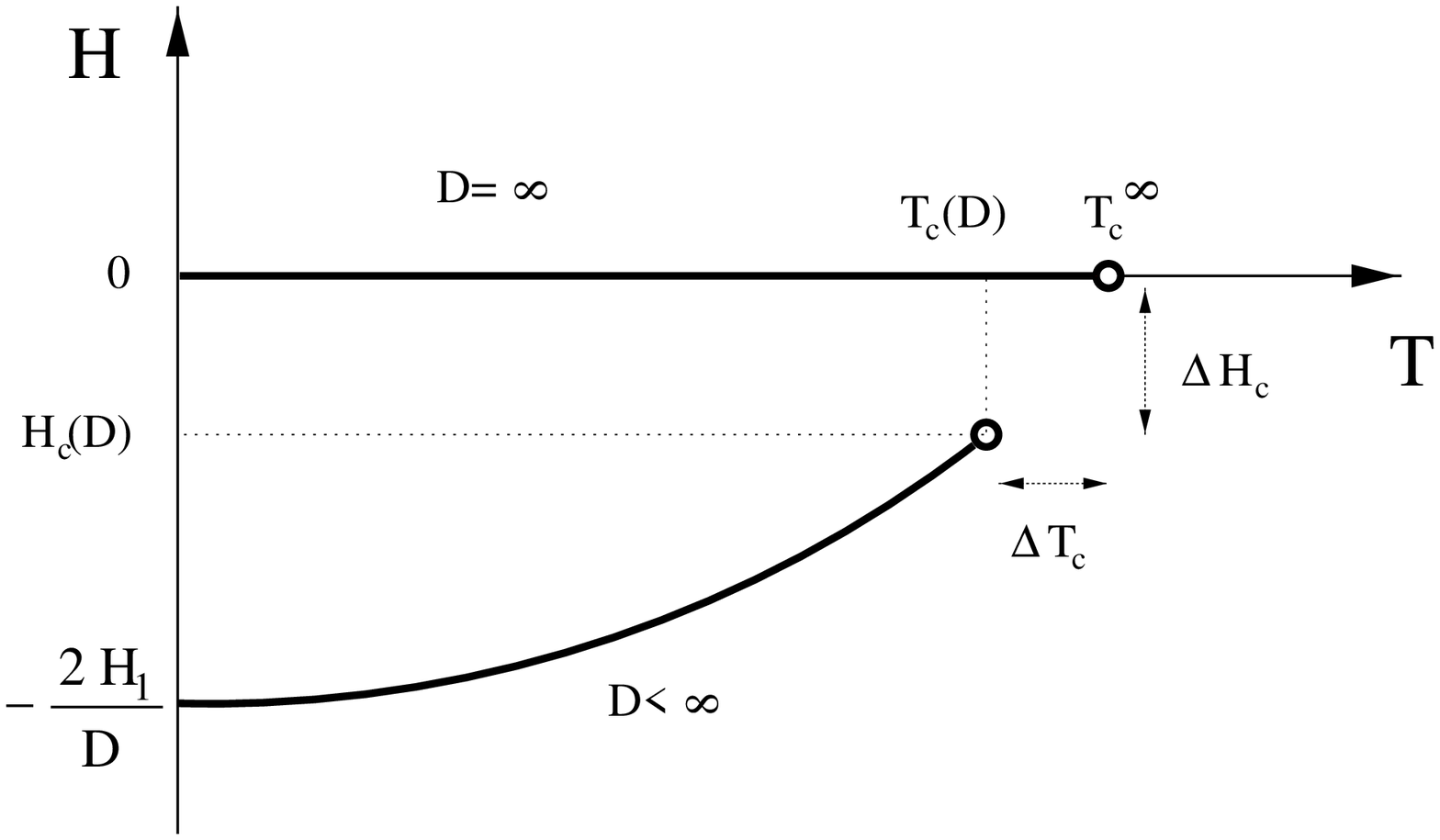,%
            clip=, angle=-0, width=120mm}
    \parbox{140mm}{\bf
	Fig. 1
    }
  \end{center}
\end{figure}

\clearpage

\begin{figure} [hbt]
  \begin{center}
    \epsfig{file=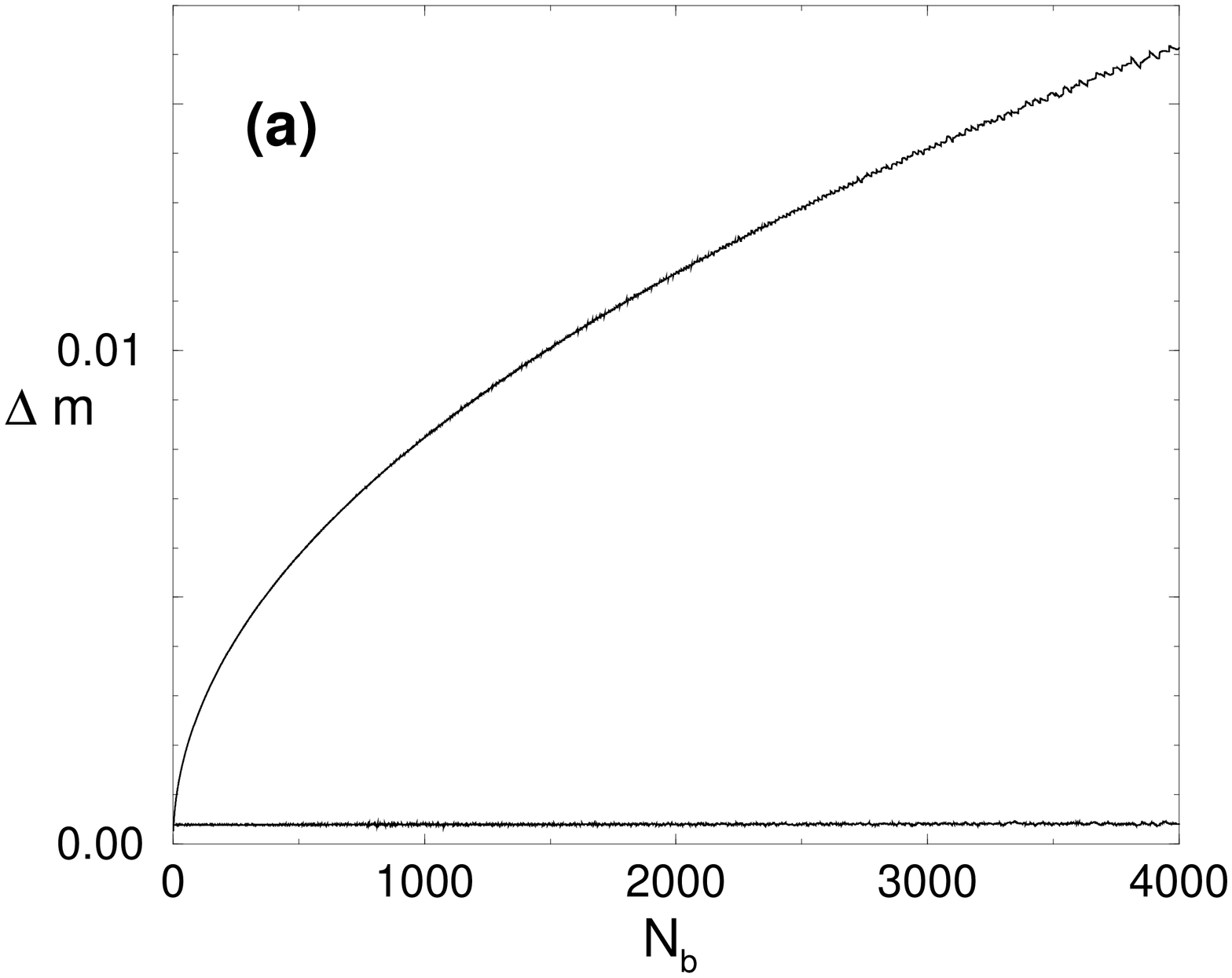,%
            clip=, angle=-90, width=100mm}
    \parbox{140mm}{\bf
	Fig. 2a
    }
  \end{center}
\end{figure}

\begin{figure} [hbt]
  \begin{center}
    \epsfig{file=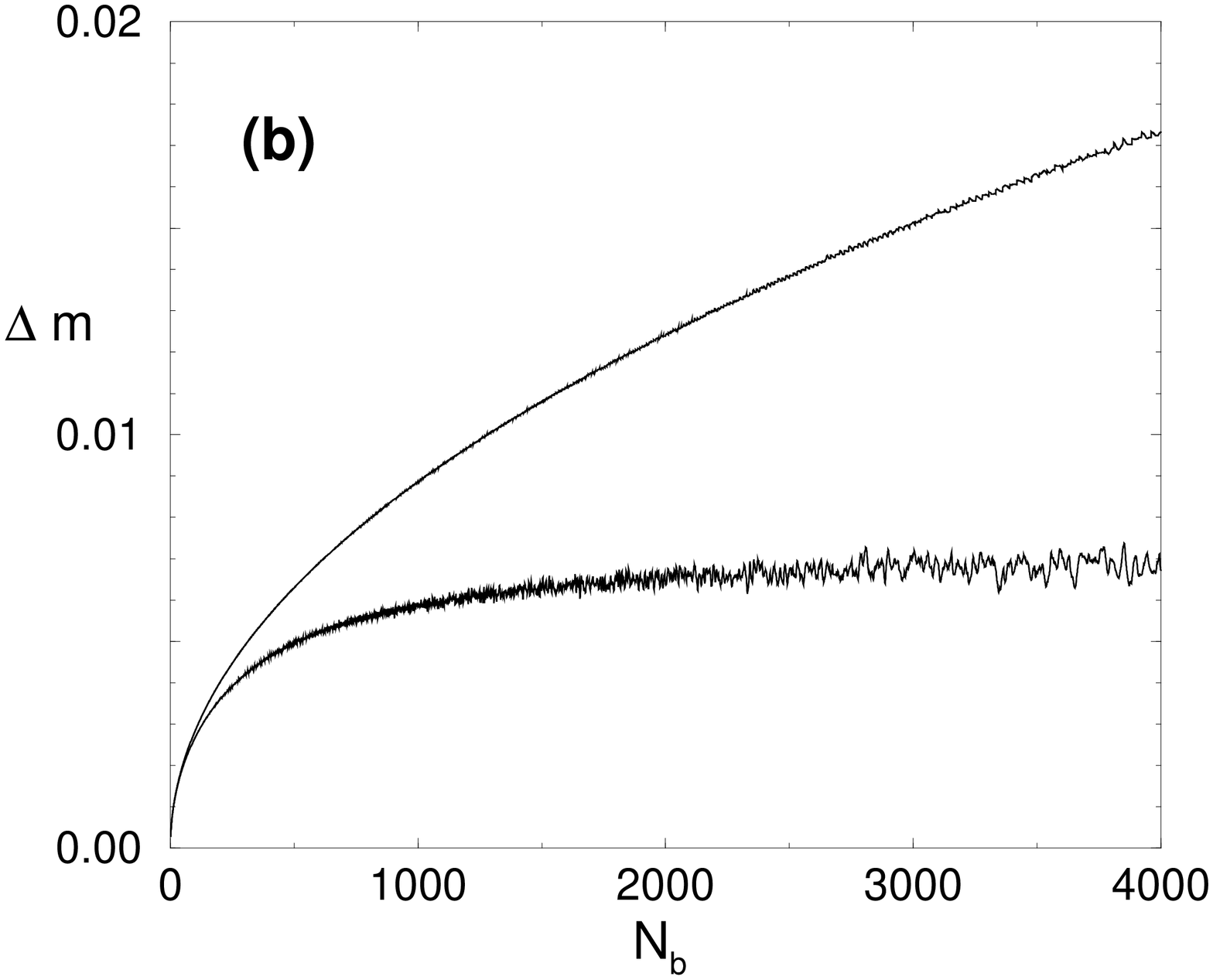,%
            clip=, angle=-90, width=100mm}
    \parbox{140mm}{\bf
	Fig. 2b
    }
  \end{center}
\end{figure}

\clearpage


\begin{figure} [hbt]
  \begin{center}
    \epsfig{file=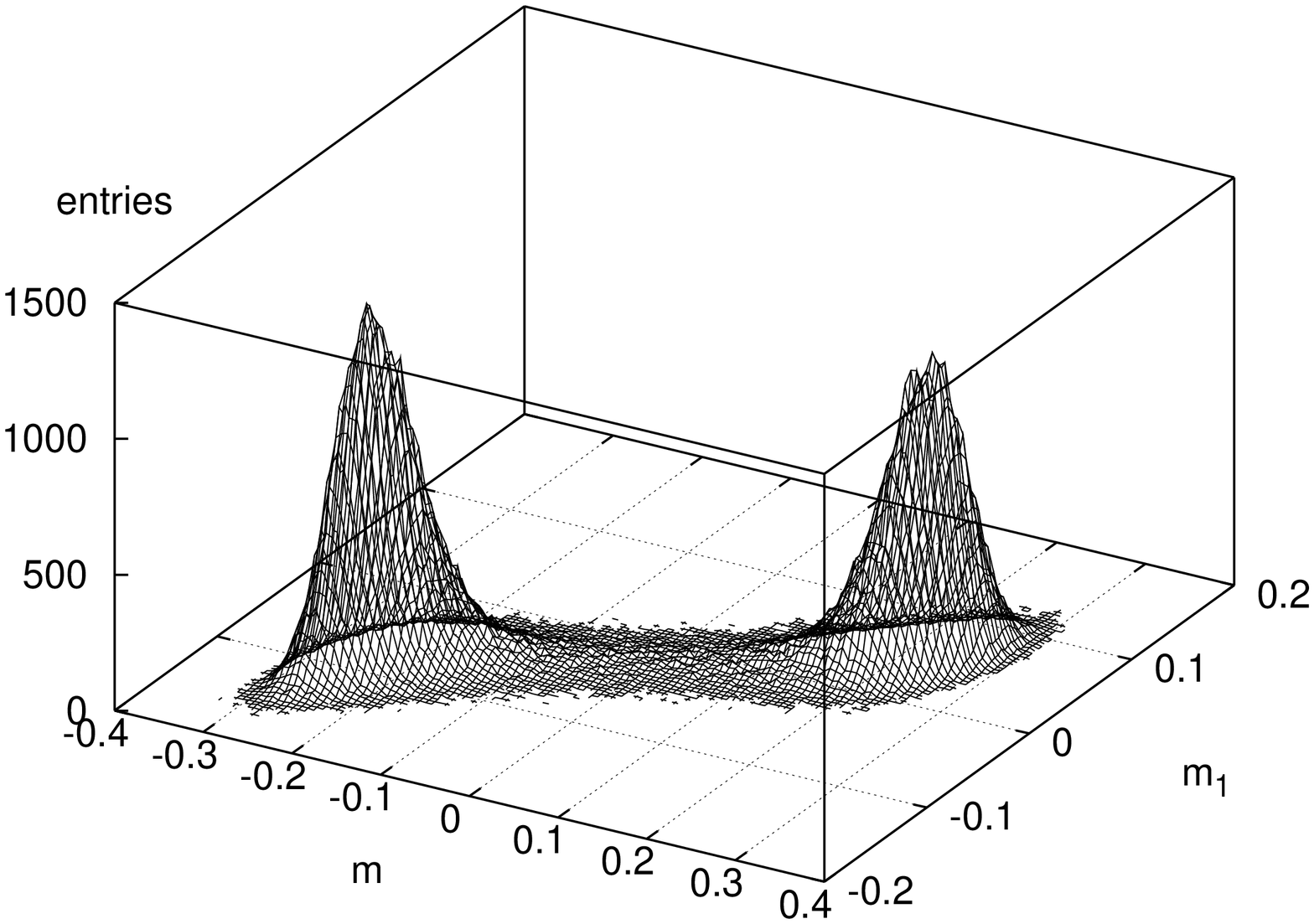,%
            clip=, angle=-90, width=170mm}
    \parbox{140mm}{\bf
	Fig. 3a
    }
  \end{center}
\end{figure}

\begin{figure} [hbt]
  \begin{center}
    \epsfig{file=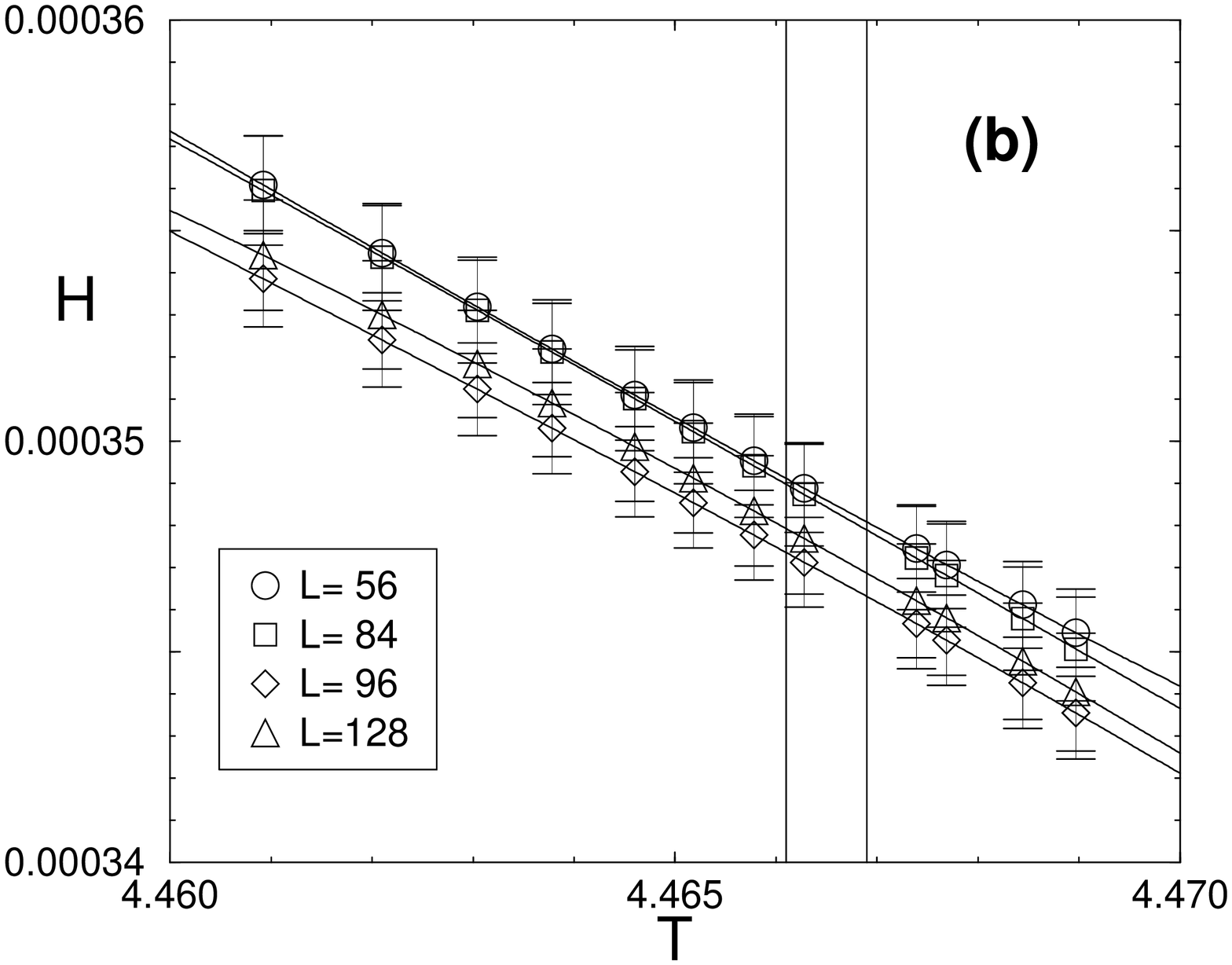,%
            clip=, angle=-90, width=120mm}
    \parbox{140mm}{\bf
	Fig. 3b
    }
  \end{center}
\end{figure}


\begin{figure} [hbt]
  \begin{center}
    \epsfig{file=fig4a.eps,%
            clip=, angle=-0, width=140mm}
    \parbox{140mm}{\bf
	Fig. 4a
    }
  \end{center}
\end{figure}

\begin{figure} [hbt]
  \begin{center}
    \epsfig{file=fig4b.eps,%
            clip=, angle=-0, width=140mm}
    \parbox{140mm}{\bf
	Fig. 4b
       }
  \end{center}
\end{figure}


\begin{figure} [hbt]
  \begin{center}
    \epsfig{file=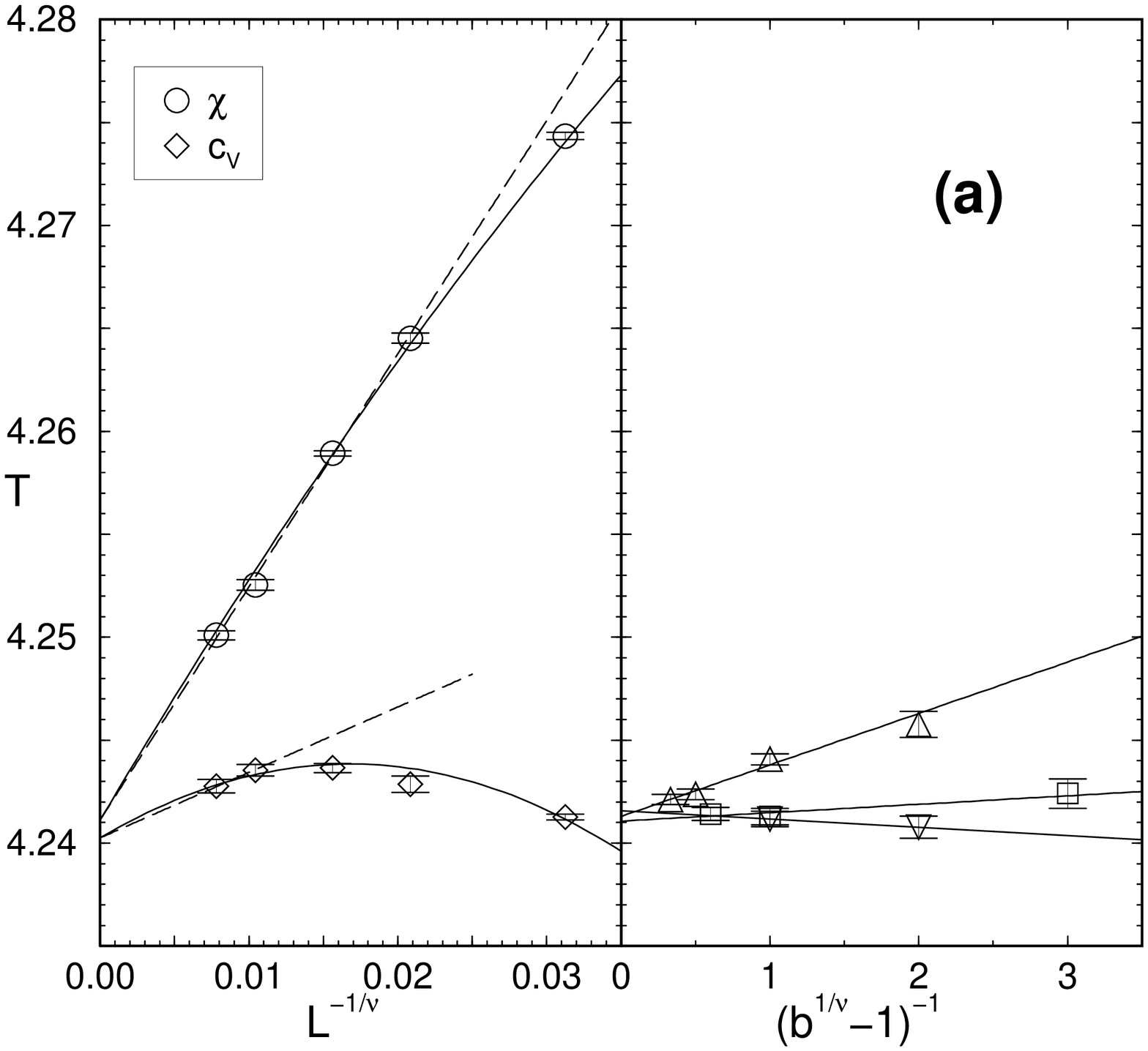,%
            clip=, angle=-90, width=120mm}
    \parbox{140mm}{\bf
	Fig. 5a
    }
  \end{center}
\end{figure}

\begin{figure} [hbt]
  \begin{center}
    \epsfig{file=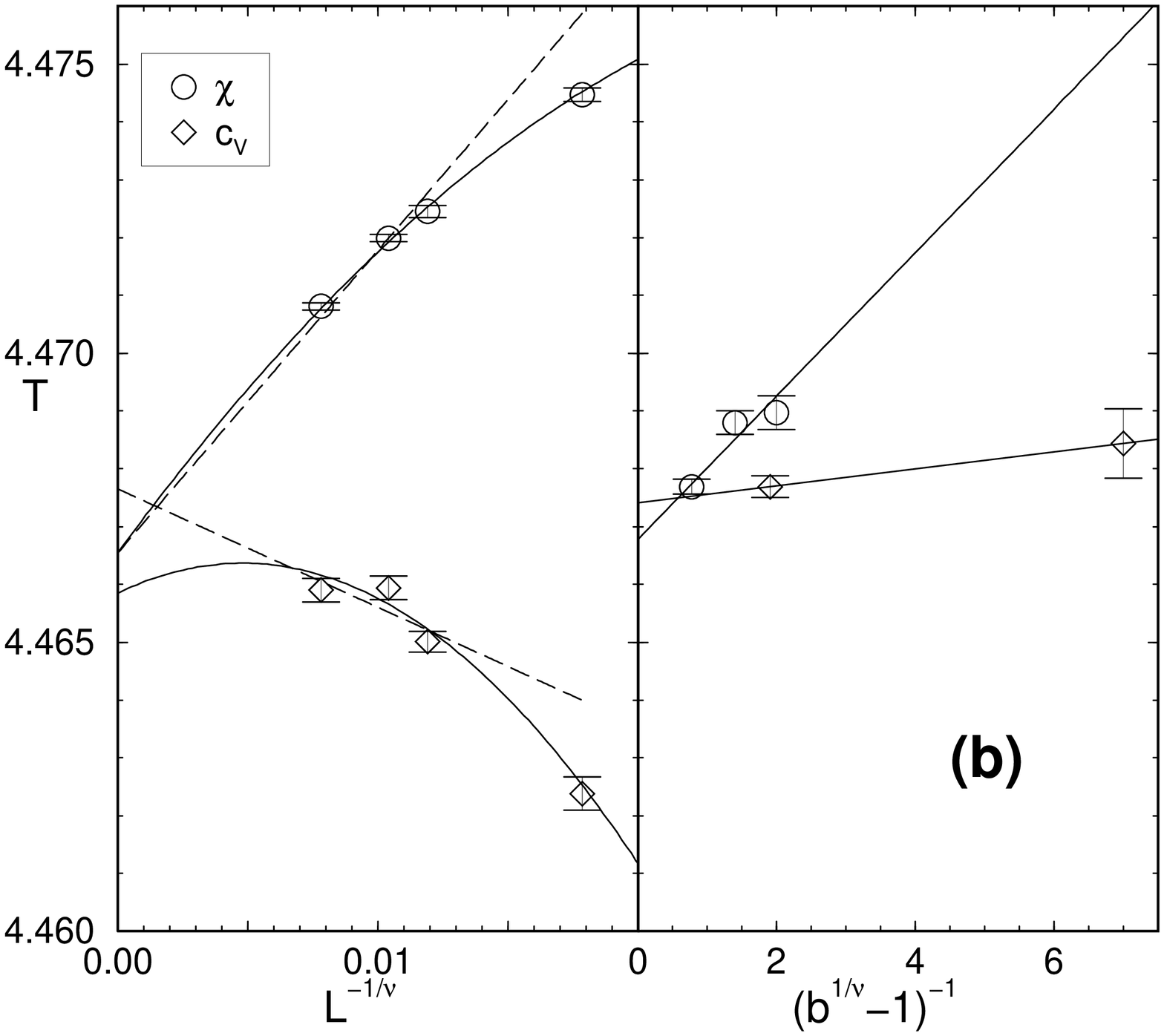,%
            clip=, angle=-90, width=120mm}
    \parbox{140mm}{\bf
	Fig. 5b
    }
  \end{center}
\end{figure}


\begin{figure} [hbt]
  \begin{center}
    \epsfig{file=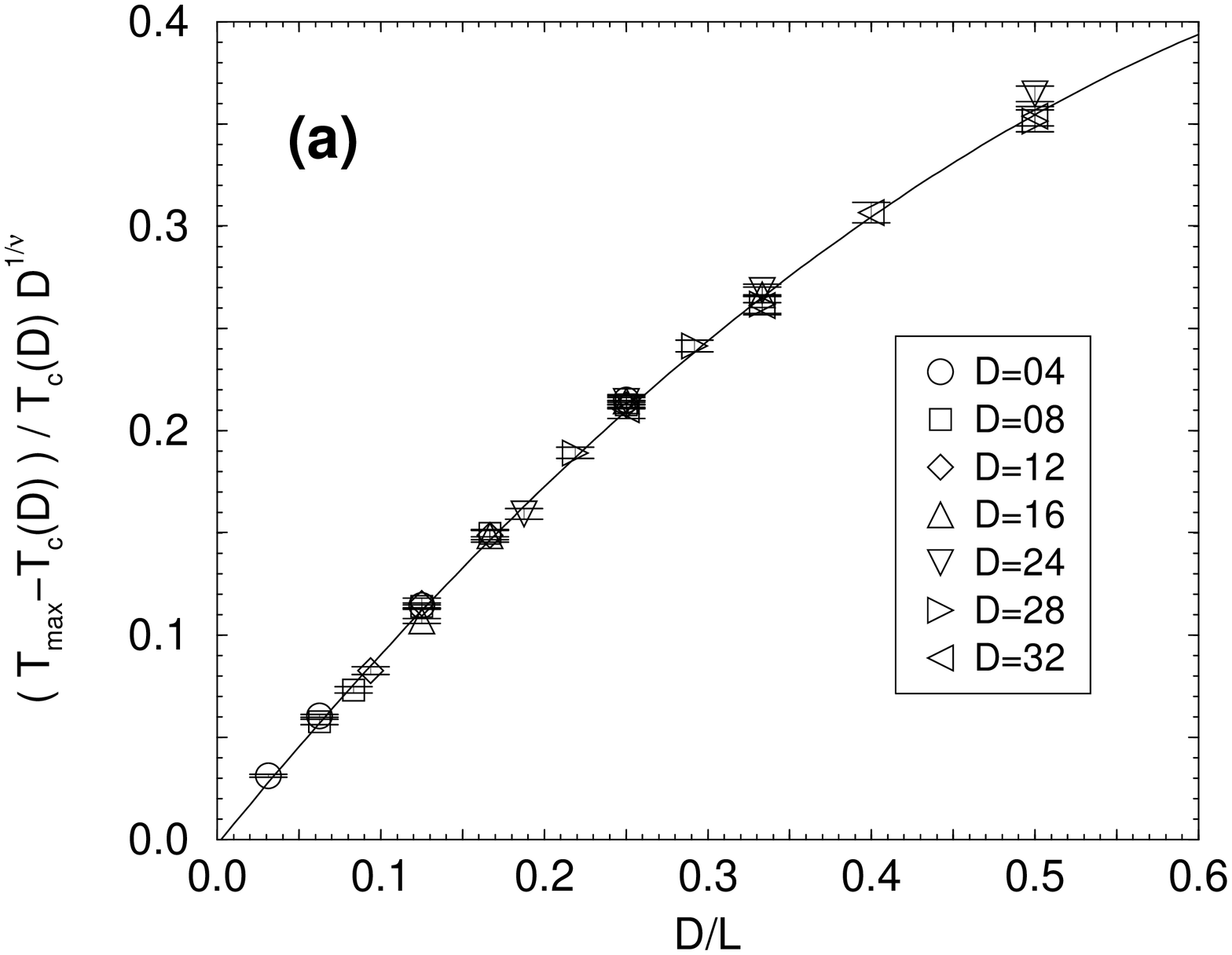,%
            angle=-90, width=120mm}
    \parbox{140mm}{\bf
	Fig. 6a
    }
  \end{center}
\end{figure}

\begin{figure} [hbt]
  \begin{center}
    \epsfig{file=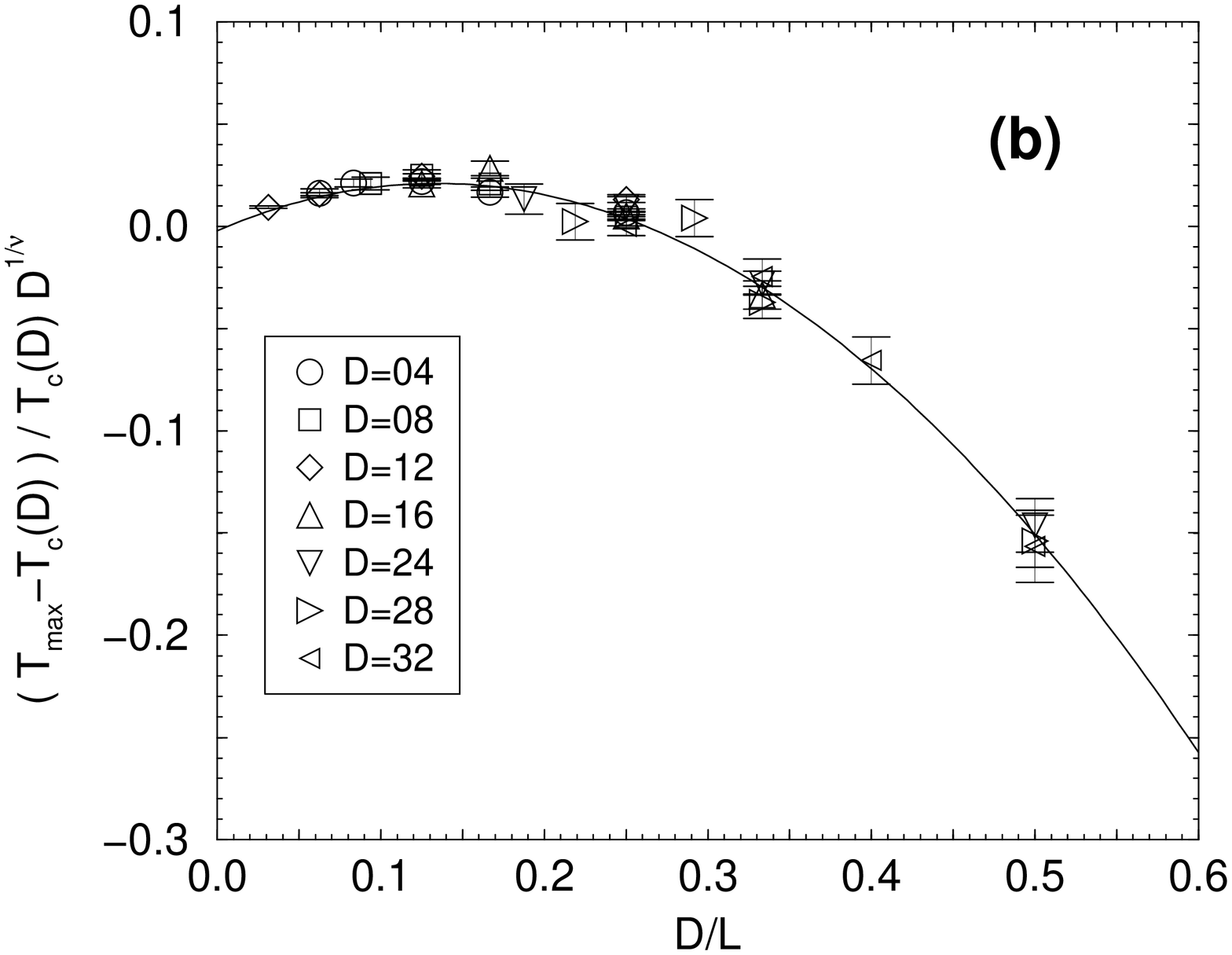,%
            angle=-90, width=120mm}
    \parbox{140mm}{\bf
	Fig. 6b
    }
  \end{center}
\end{figure}


\begin{figure} [hbt]
  \begin{center}
    \epsfig{file=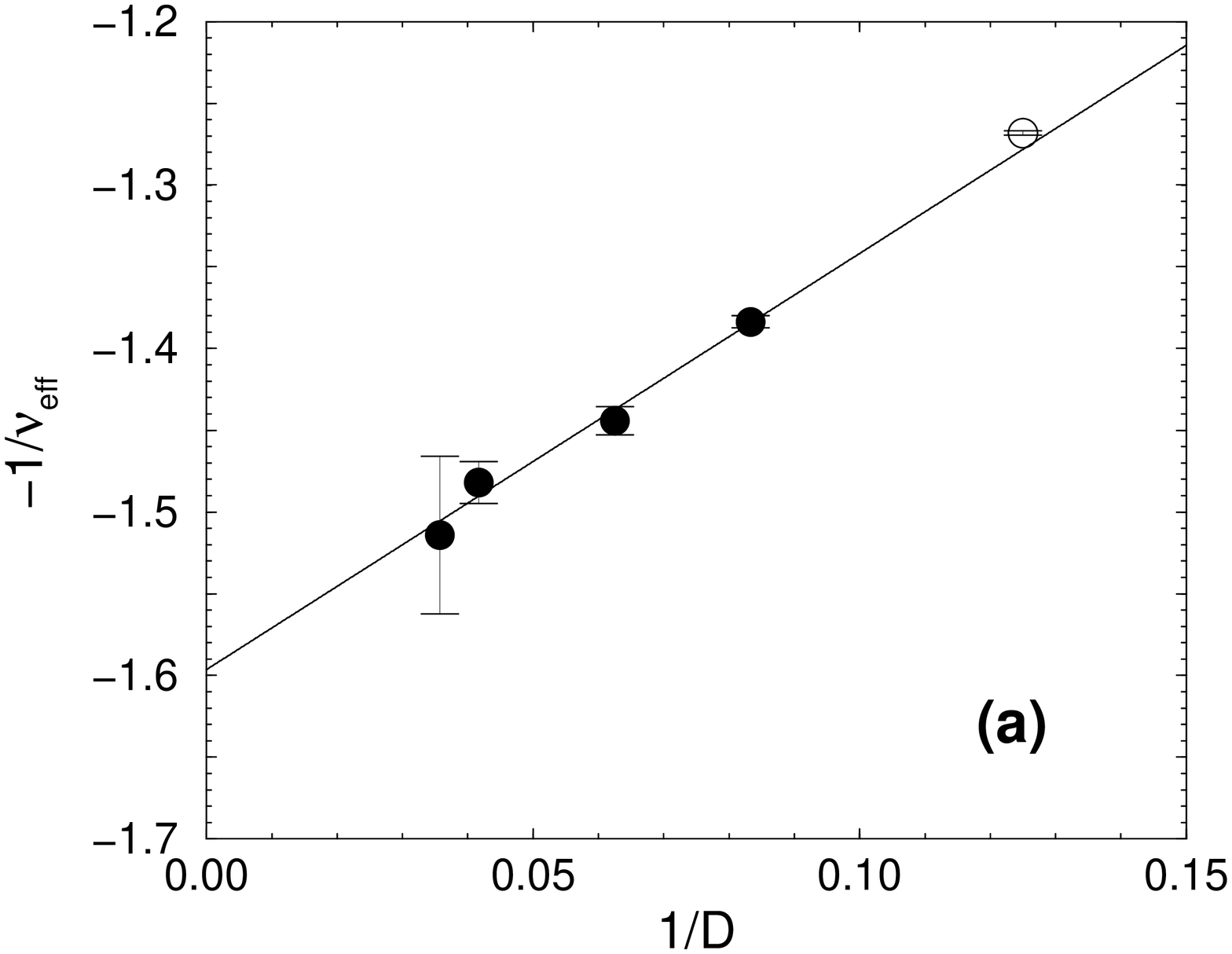,%
            clip=, angle=-90, width=140mm}
    \parbox{140mm}{\bf
	Fig. 7a
    }
  \end{center}
\end{figure}

\begin{figure} [hbt]
  \begin{center}
    \epsfig{file=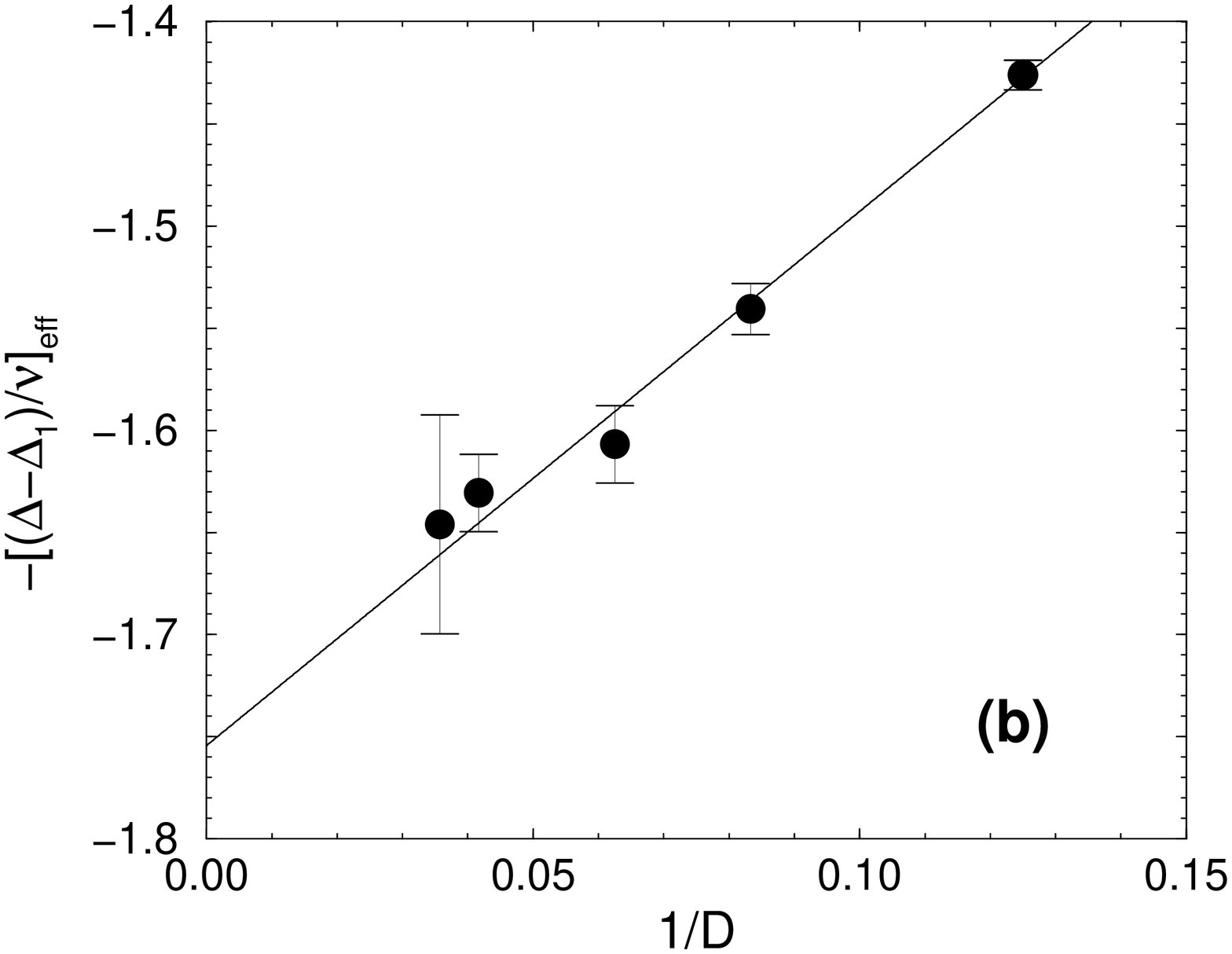,%
            clip=, angle=-90, width=140mm}
    \parbox{140mm}{\bf
	Fig. 7b
       }
  \end{center}
\end{figure}


\begin{figure} [hbt]
  \begin{center}
    \epsfig{file=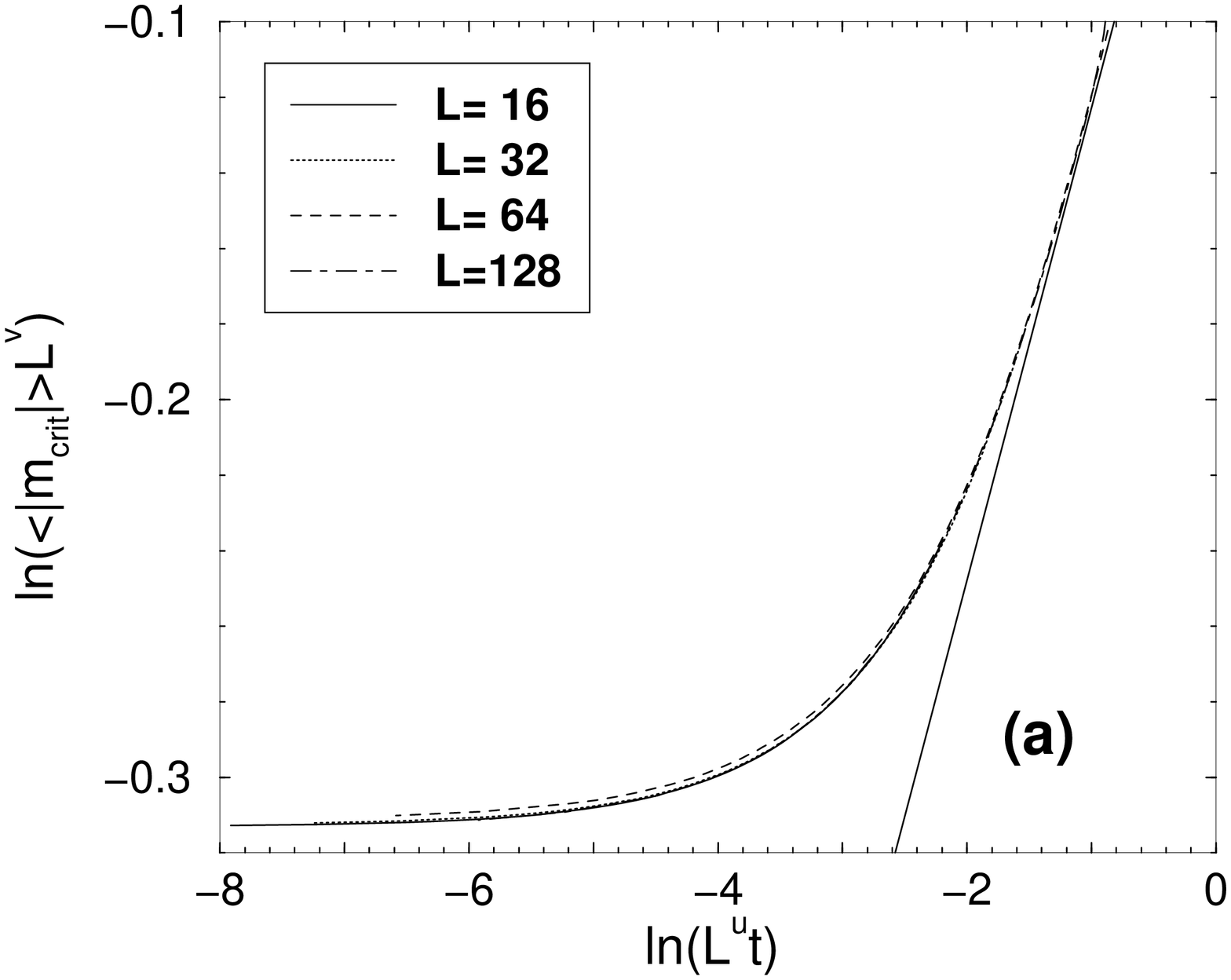,%
            angle=-90, width=90mm}
    \parbox{140mm}{\bf
	Fig. 8a
    }
  \end{center}
\end{figure}

\begin{figure} [hbt]
  \begin{center}
    \epsfig{file=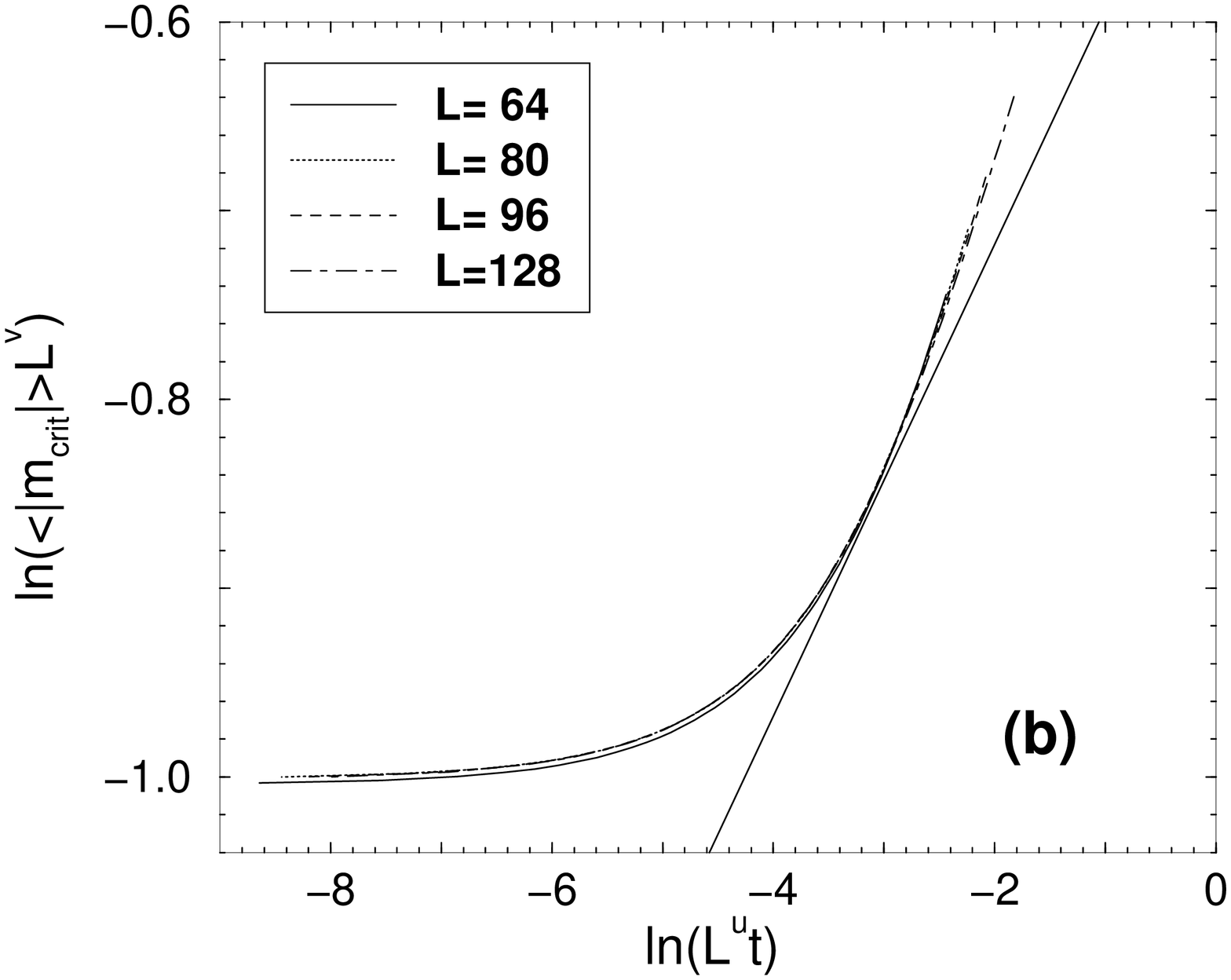,%
            angle=-90, width=90mm}
    \parbox{140mm}{\bf
	Fig. 8b
    }
  \end{center}
\end{figure}

\begin{figure} [hbt]
  \begin{center}
    \epsfig{file=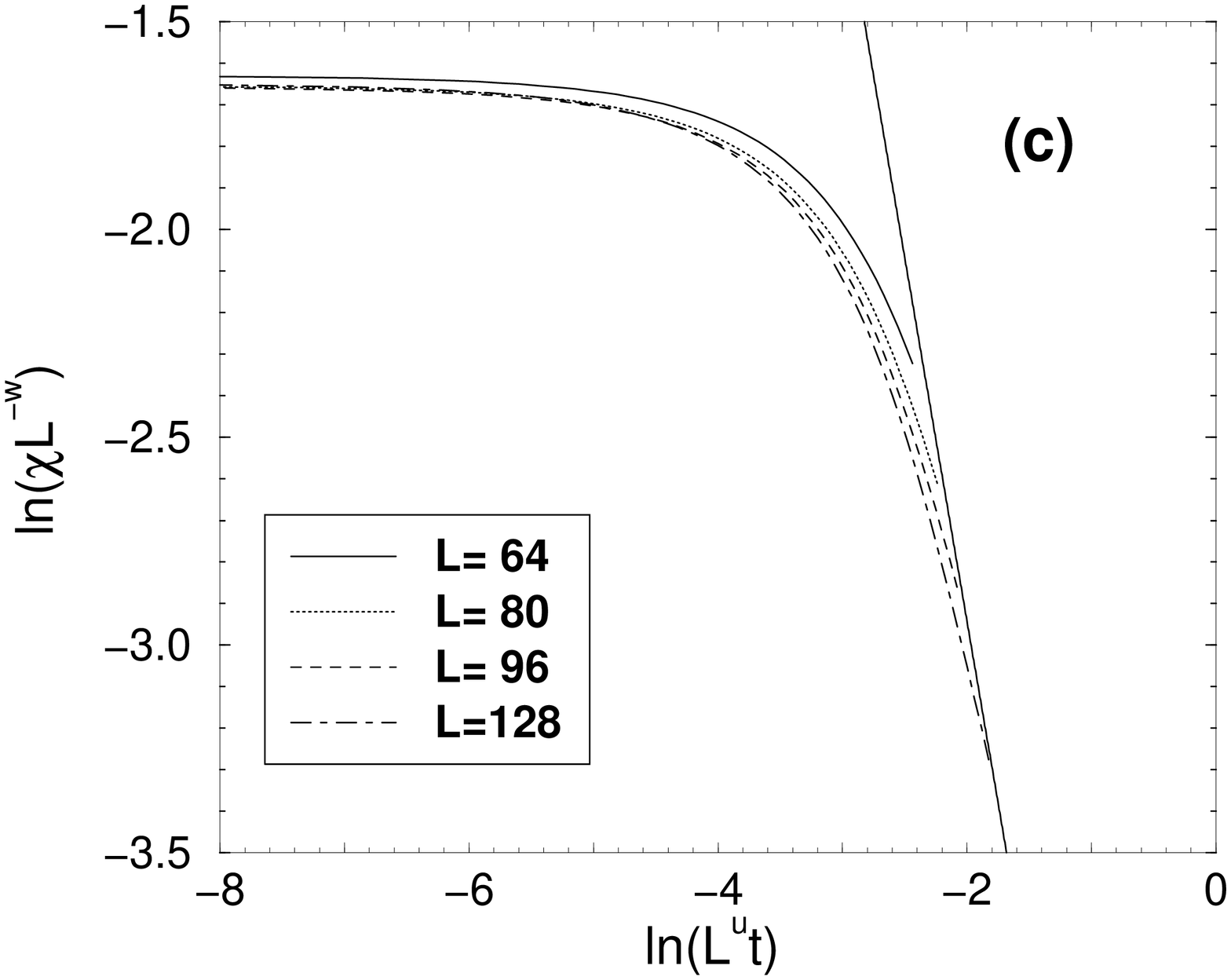,%
            angle=-90, width=90mm}
    \parbox{140mm}{\bf
	Fig. 8c
    }
  \end{center}
\end{figure}

\end{document}